\documentclass[12pt]{iopart}
\usepackage{iopams}        
\usepackage{graphicx}
\usepackage{amssymb}
\usepackage{tikz}
\usepackage{amsfonts}
\usepackage{overunderset}
\newcounter{storeeqn}
\numbysectrue
\def\be{\begin{equation}}
\def\ee{\end{equation}}
\def\ba{\begin{eqnarray}}
\def\ea{\end{eqnarray}}
\def\bigboxit#1#2{\hbox{\lower#1pt\vbox{\hrule\hbox{\vrule\kern3pt
  \vbox{\kern3pt\hbox{#2}\kern3pt}\kern3pt\vrule}\hrule}}}

\def\halfs{{\scriptstyle{\frac 12}}}
\def\Btau2{{\boldsymbol \tau}^{\vphantom{'}}_{2}}
\def\btau{{\boldsymbol \tau}^{\vphantom{'}}}
\newfont{\mycal}{eufb10 at 12pt}
\newfont{\myeu}{eurm10 at 12 pt}
\newfont{\bfrak}{eufb10 at 12 pt}
\def\rq{{\rm{q}}}

\def\mod{{\rm{mod}}}
\def\bZ{{\bf Z}}
\def\bX{{\bf X}}

\def\bQ{{\bf Q}}
\def\sfactor#1#2{\Bigg[\begin{array}{@{}c@{}}#1\\#2\end{array}\Bigg]}
\def\bino#1#2{\Bigg(\begin{array}{@{}c@{}}#1\\#2\end{array}\Bigg)}
\def\vp{^{\vphantom{.}}}
\newcount\mya\newcount\myb\font\my=cmr10 at 15pt\def\myphi
{\hbox{\my\char'010}}
\def\hypp#1#2#3#4#5{\mya=#1\myb=#1\advance\myb by1
\,{}_{\the\myb}\myphi_{\the\mya}\left[{{#2,#3}\atop{\phantom{\omega,}#4}};
#5\right]}
\begin{document}
\title[CSOS models from chiral Potts models]%
{CSOS models descending from chiral Potts models: \\
Degeneracy of the eigenspace and loop algebra}

\author{Helen Au-Yang and Jacques H H Perk}

\address{Department of Physics, Oklahoma State University, 
145 Physical Sciences, Stillwater, OK 74078-3072, USA}
\ead{perk@okstate.edu, helenperk@yahoo.com}

\begin{abstract}
Monodromy matrices of the $\Btau2$ model are known to satisfy a Yang--Baxter equation
with a six-vertex $R$-matrix as the intertwiner. The commutation relations of the elements
of the monodromy matrices are completely determined by this $R$-matrix. We show the
reason why in the superintegrable case the eigenspace is degenerate, but not in the
general case. We then show that the eigenspaces of special CSOS models descending
from the chiral Potts model are also degenerate. The existence of an $L({\mathfrak{sl}}_2)$
quantum loop algebra (or subalgebra) in these models is established by showing that the
Serre relations hold for the generators. The highest weight polynomial (or the Drinfeld
polynomial) of the representation is obtained by using the method of Baxter for the
superintegrable case. As a byproduct, the eigenvalues of all such CSOS models are
given explicitly.
\end{abstract}

\section{Introduction}
After the discovery of the integrable chiral Potts model \cite{AMPTY}\footnote{The early
history of the discovery and study of the integrable chiral Potts model is presented in \cite{JHHP}.},
the proper parametrization of the Boltzmann weights has been established
in collaboration with Professor Baxter \cite{BPA}, who has contributed a great deal also
to the further development of the theory since that time. It seems fitting, therefore,
to present a related work in the special issue in Baxter's honor. We start with a brief discussion
of how the chiral Potts model relates in two different ways to the six-vertex
model \cite{LiebWu}.
\subsection{The construction of Bazhanov and Stroganov:
descendants of six-vertex model} 
It has been noted by Baxter that the transfer matrices of six-vertex models
commute \cite{Baxter1971} and that their Boltzmann weights satisfy
Yang--Baxter equations \cite{Baxter1972}, i.e.,
\ba\fl
\sum_{\ell_2,m_2,n_2=0}^1{\cal R}(rq)_{n_1,m_1}^{n_2,m_2}
{\cal R}(pr)_{\ell_1,n_2}^{\ell_2,n_3}{\cal R}(pq)_{\ell_2,m_2}^{\ell_3,m_3}\nonumber\\
=\sum_{\ell_2,m_2,n_2=0}^1{\cal R}(pq)_{\ell_1,m_1}^{\ell_2,m_2}
{\cal R}(pr)_{\ell_2,n_1}^{\ell_3,n_2}{\cal R}(rq)_{n_2,m_2}^{n_3,m_3},
\label{ybe6v}\ea
for $\ell_i, n_i,m_i=0,1$. The $R$-matrix ${\cal R}(rq)$ is known to be the
intertwiner of the two-dimensional representations $\pi_r$ and $\pi_q$ of the
quantum group\footnote{We follow
Jimbo's review \cite{Jimbo} here. The structure involved has been recognized
as a Hopf algebra \cite{Drinfeld1985,Jimbo1985,Jimbo1986,Sklyanin}, for which
the term `quantum group' was first coined by Drin'feld \cite{Drinfeld1987}.}
${\rm U}_{\rm q}({\widehat{\mathfrak{sl}}}_2)$.
Bazhanov and Stroganov \cite{BS} found a $2\times N$ $L$-operator, with $N$ odd,
satisfying the Yang--Baxter equations
\ba\fl
\sum_{\alpha_2=0}^{N-1}\sum_{m_2,n_2=0}^1{\cal R}(rq)_{n_1,m_1}^{n_2,m_2}
{\cal L}(pp'r)_{\alpha_1,n_2}^{\alpha_2,n_3}
{\cal L}(pp'q)_{\alpha_2,m_2}^{\alpha_3,m_3}\nonumber\\
=\sum_{\alpha_2=0}^{N-1}\sum_{m_2,n_2=0}^1
{\cal L}(pp'q)_{\alpha_1,m_1}^{\alpha_2,m_2}{\cal L}(pp'r)_{\alpha_2,n_1}^{\alpha_3,n_2}
{\cal R}(rq)_{n_2,m_2}^{n_3,m_3},\nonumber\\
\quad(\alpha_1,\alpha_3=0,\cdots,N-1,\quad m_1,n_1,m_3,n_3=0,1).
\label{ybeLL}\ea
The $L$-operator intertwines a cyclic and a spin-$\frac12$ representation of
${\rm U}_{\rm q}({\mathfrak{sl}}_2)$ \cite{Jimbo}. Bazhanov and Stroganov \cite{BS} finished
their construction by recognizing that the square of four $N$-state chiral-Potts
Boltzmann weights with $N$ odd intertwines two cyclic representations.
\subsection{Six-vertex and $\tau_2$ model descending from chiral Potts model}
Not satisfied with a construction valid only for $N$ odd, in \cite{BBP}
the authors consider a square weight given by
\begin{figure}
\begin{center}
\includegraphics[width=1.5in]{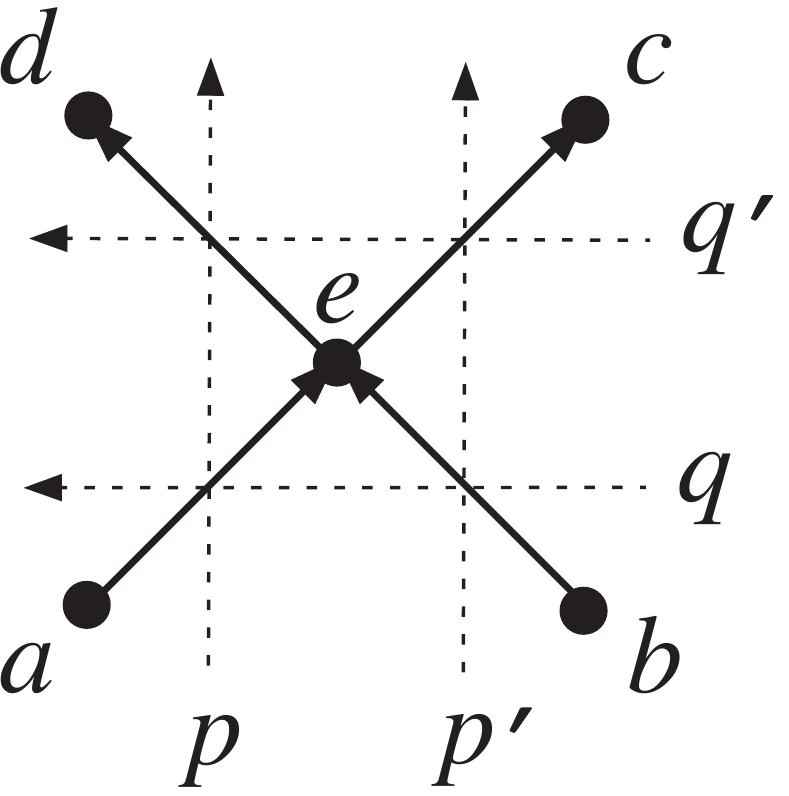}
\end{center}
\caption{The square weight $U_{pp'qq'}(a,b,c,d)$ with sum over the central spin $e$.
When $(x\vp_{q'},y\vp_{q'},\mu\vp_{q'})=(y\vp_q,\omega^\ell x\vp_q,\mu_q^{-1})$,
this can be arranged as a block-triangular array with $\ell\times\ell$ and
$(N\!-\!\ell)\times(N\!-\!\ell)$ diagonal blocks $U_{pp'q}^{(\ell)}(a,b,c,d)$ and
$U_{pp'q}^{(N-\ell)}(a,b,c,d)$ of $N\!\times\!N$ matrices.
When also $(x\vp_{p'},y\vp_{p'},\mu\vp_{p'})=(y\vp_p,\omega^j x\vp_p,\mu_p^{-1})$, 
these blocks further triangularize with diagonal blocks including the $j\times j$ block
$U_{pq}^{(\ell,j)}(a,b,c,d)$. Setting $\ell=j=2$ this block gives six-vertex model
weights at an $N$th root of unity. For general $\ell$ and $j$ the
$U_{pq}^{(\ell,j)}(a,b,c,d)$ is a Boltzmann weight of a cyclic solid-on-solid
(CSOS) model.}
\label{fig1}\end{figure}
\be\fl
U_{pp'qq'}(a,b,c,d)=\sum_{e=1}^N W_{pq}(a-e){\overline W}_{pq'}(e-d)
{\overline W}_{p'q}(b-e)W_{p'q'}(e-c),
\label{square}\ee
see figure \ref{fig1}, in which the four chiral Potts weights are given by
\be\fl
W_{pq}(n)=\Big({\mu_p\over\mu_q}\Big)^{n}\prod_{m=1}^n
\frac{y_q-x_p\omega^m}{ y_p-x_q\omega^m},\quad
  \overline W_{pq}(n)=\big({\mu_p\mu_q}\big)^{n}\prod_{m=1}^n
\frac{\omega x_p-x_q\omega^m}{ y_q-y_p\omega^m},
\label{weights}\ee
where the weights are periodic functions of $N$, $W(n+N)=W(n)$ and $\omega^N=1$. 
The subscripts $p$ and $q$ denote points on a high-genus curve, with each point $p$
parametrized by the triple $(x_p,y_p,\mu_p)$ restricted by the conditions
\ba
k'{}^2+k^2=1,\quad
k y_p^N=1-k'\mu_p^N,\quad k x_p^N=1-k'/\mu_p^N,\nonumber\\
t_p\equiv x_p y_p,\quad k^2 t_p^N=1+k'{}^2-k'(\lambda\vp_p+\lambda^{-1}_p),
\quad \lambda\vp_p\equiv\mu^N_p.
\label{xyt}\ea
In \cite{BBP} they find that when
$(x\vp_{q'},y\vp_{q'},\mu\vp_{q'})=(y\vp_q,\omega^\ell x\vp_q,\mu_q^{-1})$,
the square in (\ref{square}) becomes block triangular:
namely $U_{pp'qq'}(a,b,c,d)=0$ for $0\le a-d\le \ell-1$ and $\ell\le b-c\le N-1$.
Now the diagonal blocks depend on the
variable $t_q=x_qy_q$ only which no longer has to lie on a high-genus curve.
Particularly, for $\ell=2$, the $2\times2$ diagonal block
$U_{pp'q}^{(2)}(a,b,c,d)$ is related to $L$-operators like in (\ref{ybeLL}).

If the vertical rapidities $p$ and $p'$ are also related by  
$(x\vp_{p'},y\vp_{p'},\mu\vp_{p'})=(y\vp_p,\omega^j x\vp_p,\mu_p^{-1})$, 
the diagonal block $U_{pp'q}^{(2)}(a,b,c,d)$ is further decomposed,
namely for $0\le d-c\le j-1$ and $j\le a-b\le N-1$, $U_{pp'q}^{(2)}(a,b,c,d)=0$.
Particularly for $j=2$, $U_{pp'q}^{(2)}$ is block triangular, with one of its
diagonal blocks related to a six-vertex model. The Yang--Baxter Equations
of the chiral Potts model split into two sets of equations in IRF
(Interaction-Round-a-Face) language as,
\ba\fl
\sum^{N}_{g=1}{U}^{(2)}_{pp'r}(a,g,e,f){U}_{pp'q}^{(2)}(b,c,g,a)
{U}^{(2,2)}_{rq}(c,d,e,g)\nonumber\\
=\sum^{N}_{g=1}{U}_{rq}^{(2,2)}(b,g,f,a){U}^{(2)}_{pp'q}(g,d,e,f)
{U}^{(2)}_{pp'r}(b,c,d,g).
\label{ybeUU}\ea
shown in  figure~2. We have, as in our previous papers \cite{APsu1,APsu2,APsu4},
chosen the convention of multiplying from up to down, as seen from the above
equation and in the figure~1.
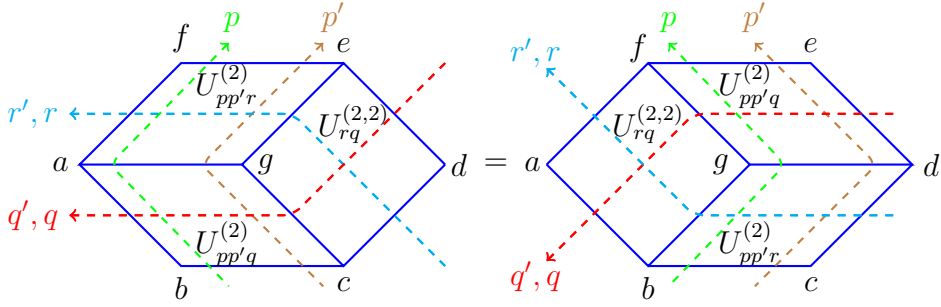
\begin{figure}
\begin{center}\hspace*{0.in}
\begin{tikzpicture}[scale=1.35]
\draw [blue,thick] (3.0,0) -- (2.0,1) -- (1.0,0) -- (2.0,-1)--(3.0,0);
\node[above] at (2.1,0.1) {$ U^{(2,2)}_{rq}$};
\draw [blue,thick]  (2.0,1) -- (0.4,1) -- (-0.6,0)--(1.,0);
\node[above] at (0.85,0.45) {$ U^{(2)}_{pp'r}$};
\draw [blue,thick] (2.0,-1) -- (0.4,-1) -- (-0.6,0);
\node [below]at (0.85,-0.45) {$ U^{(2)}_{pp'q}$};
\node at (3.5,0) {$=$};

\draw [blue,thick] (4,0) -- (5,1) -- (6,0) -- (5,-1)--(4,0);
\node [above]at (5,0.1) {$ U^{(2,2)}_{rq}$};
\draw [blue,thick]  (5,1) -- (6.6,1) -- (7.6,0)--(6,0);
\node [below]at (6.0,-0.45) {$ U^{(2)}_{pp'r}$};
\draw [blue,thick]  (5,-1) -- (6.6,-1) -- (7.6,0);
\node[above] at (6.0,0.45) {$  U^{(2)}_{pp'q}$};

\node [left] at (-0.6,0) {$a$};\node [left] at (4.05,0) {$a$};
\node [below] at (0.4,-1)  {$b$};\node [below] at (5,-1)  {$b$};
\node [below] at (2.0,-1)  {$c$};\node [below] at (6.6,-1)  {$c$};
\node [right] at (2.95,0)  {$d$};\node [right] at (7.6,0)  {$d$};
\node [above] at (2.0,1)  {$e$};\node [above] at (6.6,1)  {$e$};
\node [above] at (0.4,1)  {$f$};\node [above] at (4.9,0.9)  {$f$};
\node [right] at (1.05,0)  {$g$};\node [left] at (5.9,0)  {$g$};

\draw [<-, rounded corners, dashed,thick,green] (0.87,1.2) -- (-0.3,0) -- (0.87,-1.2);
\draw [<-, rounded corners, dashed,thick,brown] (1.8,1.2) -- (.6,0) -- (1.8,-1.2);
\node [above,green] at (0.9,1.2)  {$p$};\node [above,brown] at (1.9,1.2)  {$p'$};
\draw [<-, rounded corners, dashed,thick,cyan] (-0.7,0.5) -- (1.5,0.5) -- (3,-1.0);
\draw [<-, rounded corners, dashed,thick,cyan] (-0.7,0.5) -- (1.5,0.5) -- (3,-1.0);
\draw [<-, rounded corners, dashed,thick,red] (-0.7,-0.5) -- (1.5,-0.5) -- (3,1.0);
\draw [<-, rounded corners, dashed,thick,red] (-0.7,-0.5) -- (1.5,-0.5) -- (3,1.0);
\node [left,cyan] at (-0.7,0.5)  {$r',r$};\node [left,red] at (-0.7,-0.5)  {$q',q$};

\draw [<-, rounded corners, dashed,thick,green] (5.2,1.2) -- (6.35,0) -- (5.2,-1.2);
\draw [<-, rounded corners, dashed,thick,brown] (6.05,1.2) -- (7.25,0) -- (6.05,-1.2);
\node [above,green] at (5.2,1.2)  {$p$};\node [above,brown] at (6.05,1.2)  {$p'$};
\draw [<-, rounded corners, dashed,thick,red] (4.0,-0.95) -- (5.45,0.5) -- (7.45,0.5);
\draw [<-, rounded corners, dashed,thick,red] (4.0,-0.95) -- (5.45,0.5) -- (7.45,0.5);
\draw [<-, rounded corners, dashed,thick,cyan] (4.0,0.95) -- (5.45,-0.5) -- (7.45,-0.5);
\draw [<-, rounded corners, dashed,thick,cyan] (4.0,0.95) -- (5.45,-0.5) -- (7.45,-0.5);
\node [left,cyan] at (4.25,1.1)  {$r',r$};\node [left,red] at (4.25,-1.1)  {$q',q$};
\end{tikzpicture}
\end{center}
\caption{Pictorial representation of Yang--Baxter equation (\ref{ybeUU}).}
\label{fig2}\end{figure}

The six-vertex $R$-matrix used by Bazhanov and Stroganov is different from the
one descending from the chiral Potts model, creating subtle differences in the
$\tau_2$-matrices in the two approaches. These differences are presented next.
\subsection{Comparison of $R$-matrices of \cite{BS} and \cite{BBP}}
\vglue5pt
\begin{minipage}[b]{5.5cm}
\begin{center}
Bazhanov and Stroganov\\
\begin{eqnarray}
\fl{\bf R}(x)= \left( 
\begin{array}{cccc} 
a & 0 & 0 & 0 \\ 
0 & b &  x c & 0\\ 
0 & c/x &{b} & 0\\
0 & 0 & 0 & a
\end{array} \right) 
\hfill\cr\cr
\fl a=(x\rq-x^{-1}\rq^{-1}),\quad b=(x-x^{-1})\cr
\fl c=(\rq-\rq^{-1}),\quad \rq^N=1,\quad\hbox{$N$ odd}.\cr
\nonumber\end{eqnarray}
\end{center}
\end{minipage}\hfill
\begin{minipage}[b]{8cm}
\begin{center}
Descendant of Chiral Potts\\
\begin{eqnarray}
\fl U^{(2,2)}_{pq}= \left( 
\begin{array}{cccc} 
a & 0 & 0 & 0 \\ 
0 & b &  t c & 0\\ 
0 & c &{b/{\omega}} & 0\\
0 & 0 & 0 & a
\end{array} \right) 
\hfill\cr\cr
\fl a=(1-\omega^{-1} t),\quad x^2=t=t_q/t_p,\quad \omega^N=1,\cr
\fl b=(1-t),\quad c=(1-\omega^{-1}),\quad \hbox{any $N$}.\cr
\label{6v}\end{eqnarray}
\end{center}
\end{minipage}
\vglue5pt
\vglue5pt
In the above comparison\footnote{We use roman ${\mathrm q}$ to distinguish
it from the rapidity variable $q$. Note that only for $N$ odd we can have
${\mathrm q}^2=\omega$ while both $\omega^N=1$ and ${\mathrm q}^N=1$.}
$\omega=\rq^2$.
The transfer matrices for the symmetric six-vertex case on the left and the
asymmetric one on the right are given respectively by
\be\fl
[T_{6v}]_{n,n'}={\rm tr}\Bigg[\prod_{i=1}^L\sum_{m_i=0}^1
{\bf R}_{n_i,m_i}^{n'_{i},m_{i+1}}(t)\Bigg],
\quad [T_{a6v}]_{\sigma,\sigma'}={\rm tr}\Bigg[\prod_{j=1}^L
{\bf U}_{r,q}^{(2,2)}(\sigma\vp_j,\sigma\vp_{j+1},\sigma'_{j+1},\sigma'_{j})\Bigg].\ee
Using Baxter's well-known method---see e.g.\ chapter 10.14 of his book
\cite{Baxterbook}---we can take the Hamiltonian limit and find
 \ba\fl T_{6v}&\to {\cal H}_{XXZ}=
\sum_{j=1}^L\big[\sigma^x_j\sigma^x_{j+1}
+\sigma^y_j\sigma^y_{j+1}+\Delta\sigma^z_j\sigma^z_{j+1}\big],
\quad\Delta=\cos \frac\pi N,\\
 \fl T_{a6v}&\to{\cal H}_{XXX}+{\cal H}_{DM}=
\sum_{j=1}^L\bigg[\Delta(\sigma^x_j\sigma^x_{j+1}
+\sigma^y_j\sigma^y_{j+1}+\sigma^z_j\sigma^z_{j+1})\nonumber\\
\fl&\hspace{1.4in}+\sin\frac\pi N (\sigma^x_j\sigma^y_{j+1}
-\sigma^y_j\sigma^x_{j+1})\bigg].
\ea
This shows that,  instead of the XXZ-spin chain Hamiltonian with periodic
boundary conditions, the asymmetric case reduces to a periodic XXX chain
with Dzyaloshinsky--Moriya term \cite{MWxxz}.\footnote{The two Hamiltonians are
related by a unitary similarity transformation \cite{PC} up to a twist in the
boundary conditions when $L$ is not a multiple of $N$.}

The $L$-operator of Bazhanov and Stroganov is
\ba
{\cal L}_j(x)= \left[ 
  \begin{array}{cc} 
x^{-1}d_-{\bf X}^{-\rho}_j+xd_+{\bf X}^{\rho}_j
&  x(g_+{\bf X}^{-\rho}_j+g_-{\bf X}^{\rho}_j){\bf Z}\vp_j \\ 
x^{-1}(h_+{\bf X}^{-\rho}_j+h_-{\bf X}^{\rho}_j) {\bf Z}_j^{-1}
& x^{-1}f_-{\bf X}^{\rho}_j+xf_+{\bf X}^{-\rho}_j
 \end{array} \right],\cr\cr
 h_+=d_-f_+/g_+,\quad h_-=d_+f_-/g_-,\quad \rho=(N-1)/2.
\label{L2}\ea
and the corresponding $\Btau2(x)$ is given by
\be
\Btau2(x)={\rm tr}[{\cal \bf L}(x)],\quad
{\cal \bf L}(x)\equiv{\cal L}_L(x)\cdots{\cal L}_j(x)\cdots{\cal L}_1(x),
\quad x^2=t_q.
\label{tau2bs}\ee
However, $\Btau2(t_q)$ descending from the chiral Potts model is
\be
\Btau2(t_q)={\rm tr}[{\boldsymbol {\cal U}}(t_q)],\quad
{\boldsymbol {\cal U}}(t_q)\equiv\prod_{i=1}^L
{\bf U}_{p^{\vp}_ip'_i,q}^{(2)}(\sigma\vp_i,\sigma\vp_{i+1},\sigma'_{i+1},\sigma'_{i}),
\label{gtau2}\ee
with $0\le\sigma_i-\sigma'_i\le 1$ for $1\le i\le L+1$,
without imposing the periodic boundary condition,
where
\ba\fl
{\bf  U}^{(2)}_{pp'q}=\left[ 
  \begin{array}{cc} 
1-\omega {t_q}{\bf Z}/y_p y_{p'}&{-\omega t_q}\mu_{p'}
(1- {x_{p'}}{\bf Z}/{y_p}){\bf X} /{y_{p'}}
 \\ 
{\mu_{p}}{\bf X}^{-1}(1-{\bf Z}  {x_{p}}/{y_{p'}})/{y_{p}}
&  {\mu_p\mu_{p'}}(\omega x_p x_{p'}{\bf Z}- \omega t_q)/{y_p y_{p'}}
 \end{array} \right].
 \label{u2}\ea
In (\ref{L2}) and  (\ref{u2}), we have
\be \fl
{\bf Z}= \left( 
  \begin{array}{ccccc} 
1&0&0&\cdots&0 \\ 
 0&\omega&0&\cdots&0\\
 0&0&\omega^2&\cdots&0\\
 \vdots&\ddots&\ddots&\ddots&\vdots\\
 0&\cdots&0&0&\omega^{N-1}
 \end{array} \right),
\quad
   {\bf X}= \left( 
  \begin{array}{ccccc} 
0&0&0&\cdots&1 \\ 
 1&0&0&\cdots&0\\
 0&1&0&\cdots&0\\
 \vdots&\ddots&\ddots&\ddots&\vdots\\
 0&\cdots&0&1&0
 \end{array} \right),\ee
 and
 \ba
  {\bf X}_j=\hbox{\vbox{\hbox{$_1$}\hbox{${\bf 1}$}}}\otimes
\hbox{\vbox{\hbox{\;$_{\cdots}$}\hbox{$\cdots$}}}\otimes{\bf 1}\otimes
\hbox{\vbox{\hbox{\;$_j$}\hbox{${\bf X}$}}}\otimes
{\bf 1}\otimes\hbox{\vbox{\hbox{\;$_{\cdots}$}\hbox{$\cdots$}}}
\otimes\hbox{\vbox{\hbox{$_L$}\hbox{${\bf 1}$}}}\,,\nonumber\\
{\bf Z}_j=\hbox{\vbox{\hbox{$_1$}\hbox{${\bf 1}$}}}\otimes
\hbox{\vbox{\hbox{\;$_{\cdots}$}\hbox{$\cdots$}}}\otimes{\bf 1}\otimes
\hbox{\vbox{\hbox{\;$_j$}\hbox{${\bf Z}$}}}\otimes
{\bf 1}\otimes\hbox{\vbox{\hbox{\;$_{\cdots}$}\hbox{$\cdots$}}}
\otimes\hbox{\vbox{\hbox{$_L$}\hbox{${\bf 1}$}}}\,.
\ea
It is worthwhile to note that, even though we use the same
${\bf Z}$ and ${\bf X}$ in both cases, in (\ref{L2}) the matrices act on
spin variables $\sigma_i$, but in (\ref{u2}) they act on edge variables
$n_i=\sigma_i-\sigma_{i+1}$ mod $N$, such that
${\bf Z}_j|\{n_i\}\rangle=\omega^{n_i}|\{n_i\}\rangle$ and
${\bf X}_j|\{n_i\}\rangle=|n_1,\cdots,n_i+1,\cdots,n_L\rangle$.

The monodromy matrix\footnote{This concept was introduced in
the quantum inverse scattering method (QISM) \cite{STF}.} in (\ref{tau2bs}),
\be
{\bf L}(x)=
\left[\begin{array}{cc}
{\bf A}(x) &{\bf B}(x)\cr
{\bf C}(x) &{\bf D}(x)\end{array}\right]=\sum_{n=-L}^L\,x^n
\left[\begin{array}{cc}
{\bf A}_n &{\bf B}_n\cr
{\bf C}_n &{\bf D}_n\end{array}\right],
\label{ABCDBS}\ee
satisfies the Yang--Baxter equation 
\be
[{\cal \bf L}(x)\otimes{\cal \bf L}(y)]{\bf {\check R}}(x/y)
={\bf {\check R}}(x/y)[{\cal \bf L}(y)\otimes{\cal \bf L}(x)].
\label{YBErLL}\ee
It is easy to show that ${\bf A}_n$ and ${\bf D}_n$ are nonvanishing
only for even $n$, while ${\bf B}_n$ and ${\bf C}_n$ are nonzero for odd $n$.
On the other hand, the monodromy matrix in (\ref{gtau2}) is 
 \be
{\boldsymbol {\cal U}}(t_q)=
\left[\begin{array}{cc}
{\bf A}(t_q) &{\bf B}(t_q)\cr
{\bf C}(t_q) &{\bf D}(t_q)\end{array}\right]=\sum_{n=0}^L\,(-\omega t_q)^n
\left[\begin{array}{cc}
{\bf A}_n &{\bf B}_n\cr
{\bf C}_n &{\bf D}_n\end{array}\right], 
\label{ABCD}\ee
with nonnegative indices $n$ for the coefficients and ${\bf B}_0={\bf C}_L=0$.
From repeated application of the Yang--Baxter equation (\ref{ybeUU}) one can
show that a similar Yang--Baxter equation holds for this monodromy matrix,
\be
[{\boldsymbol {\cal U}}(t_r)\otimes{\boldsymbol {\cal U}}(t_q)]{\bf U}^{(2,2)}_{rq}
={\bf U}^{(2,2)}_{rq}[{\boldsymbol {\cal U}}(t_q)\otimes{\boldsymbol {\cal U}}(t_r)].
\label{YBEuUU}\ee

Both Yang--Baxter equations (\ref{YBEuUU}) and (\ref{YBErLL}) give rise to
sixteen relations between the ${\bf A}(t)$, ${\bf B}(t)$, ${\bf C}(t)$ and ${\bf D}(t)$.
By changing the vertical rapidity variables, or changing the size (or length) $L$,
we change the monodromy matrices, but that does not change the Yang--Baxter
equations. Thus, the sixteen relations remain the same in each of the two cases. 
However, the differences in the six-vertex $R$-matrices  shown in (\ref{6v})
cause the sixteen relations to be different for the two cases (\ref{ABCDBS})
and (\ref{ABCD}).
\subsection{Degenerate eigenspace in XXZ model and superintegrable $\tau_2$ model}  
For the superintegrable chiral Potts model, it was shown in
\cite{vonGehlen1985, AMP,Baxter1988,Baxter1989,BaxterSB}
that there exist special sets of $2_{\vp}^{m_E}$ Ising-like eigenvalues
of the transfer matrix or Hamiltonian, which implies a
$2_{\vp}^{m_E}$-fold degeneracy in the corresponding $\tau_2$ model.
Superintegrability means that the model satisfies two or more different
integrability criteria, like Yang--Baxter or Onsager algebra integrability
\cite{JHHP,AMP}.
In their study of the XXZ model at roots of unity \cite{DFM, Degu1,Degu2,Degu3},
the authors show the existence of a quantum loop algebra
$L({\widehat{\mathfrak{sl}}}_2)$ in the XXZ model. Such a loop algebra or
subalgebra  was also shown in \cite{APsu1,APsu2,APsu4,NiDe1,NiDe2}
to exist in certain sectors of the superintegrable $\Btau2$-model.
The proof of this degeneracy is based on the sixteen relations
of the Yang--Baxter equations. Since the equations (\ref{ybeUU}) are
model-independent, one needs to know why there is degeneracy in the
superintegrable $\Btau2$ model, but not in the generic $\Btau2$ model. 
 \subsection{Understanding the degeneracy} 
Consider two of the sixteen equations obtained from (\ref{YBEuUU}),
\ba
(\omega y-x ){\bf A}(x){\bf B}(y)=
\omega(y-x){\bf B}(y){\bf A}(x)+y(\omega-1){\bf A}(y){\bf B}(x),
\label{AB}\\
(\omega y-x ){\bf B}(y){\bf D}(x)=
(y-x ){\bf D}(x){\bf B}(y)+
y(\omega-1){\bf B}(x){\bf D}(y).
\label{DB}\ea
Equating the coefficients of $x^{L+1}$ of these two equations, we find
\be
{\bf A}_L{\bf B}(y)=\omega{\bf B}(y){\bf A}_L,\quad
{\bf B}(y){\bf D}_L={\bf D}_L{\bf B}(y).
\ee
Similarly by equating the coefficients of $y^{L+1}$, we have
\ba
{\bf A}(x){\bf B}_L-{\bf B}_L{\bf A}(x)
=(1-\omega^{-1}){\bf A}_{L}{\bf B}(x)=-(1-\omega){\bf B}(x){\bf A}_{L},
\nonumber\\
{\bf D}(x){\bf B}_L-\omega{\bf B}_L{\bf D}(x)
=(1-\omega){\bf D}_{L}{\bf B}(x)=(1-\omega){\bf B}(x){\bf D}_{L}.
\ea
By induction, one can show
\ba
{\bf A}(x){\bf B}^n_L-{\bf B}^n_L{\bf A}(x)
=-(1-\omega)[n]{\bf B}(x){\bf B}^{n-1}_L{\bf A}_{L},
\nonumber\\
{\bf D}(x){\bf B}^n_L-\omega^n{\bf B}^{n}_L{\bf D}(x)
=(1-\omega)[n]{\bf B}(x){\bf B}^{n-1}_L{\bf D}_{L}.
\label{commABn}\ea
so that
\be\fl
[{\bf A}(x)+{\bf D}(x)]{\bf B}^{(n)}_L-{\bf B}^{(n)}_L[{\bf A}(x)+\omega^n{\bf D}(x)]
=(\omega-1){\bf B}(x){\bf B}^{(n-1)}_L[{\bf A}_{L}-{\bf D}_{L}].
\label{commABnn}\ee
Here we have used the definitions
\be {\bf B}^{(n)}_L= \frac{{\bf B}^{\,n}_L}{[n]!},
\qquad [n]=\frac {1-\omega^{n}}{1-\omega},
\quad [n]!=\prod_{k=1}^n\,[k].
\ee
Letting $n=N$ in (\ref{commABnn}), and using $\omega^N=1$, 
we find both $ {\bf B}^{\,N}_L=0$ and $[N]!=0$. However, $ {\bf B}^{(N)}_L$
can be defined through a limiting procedure \cite{NiDe2}, so that
\be\fl
[\Btau2(x),{\bf B}^{(N)}_L]=[{\bf A}(x)+{\bf D}(x),{\bf B}^{(N)}_L]
=(\omega-1){\bf B}(x){\bf B}^{(N-1)}_L[{\bf A}_{L}-{\bf D}_{L}].
\ee
This shows that the degeneracy of the eigenspace of an eigenvalue
depends on the difference ${\bf A}_{L}-{\bf D}_{L}$.
For, if $|\Psi\rangle$ is an eigenvector of $\Btau2(x)$ and
$[{\bf A}_{L}-{\bf D}_{L}]|\Psi\rangle=0$, then 
$[\Btau2(x),{\bf B}^{(N)}_L]|\Psi\rangle=0$.
Consequently, ${\bf B}^{(N)}_L|\Psi\rangle$ is also an eigenvector
with same eigenvalue. 

For the generic case, we can show that ${\bf B}^N_L\ne0$,
but constant, so that ${\bf B}^{N}_L|\Psi\rangle$ does not give
rise to an independent eigenvector, and its eigenspace is nondegenerate.

From (\ref{u2}) we have
\ba
{\bf A}_0=1,
&&{\bf D}_0=\omega^L \prod_{j=1}^L\frac{{\bf Z}_j\mu_{p\vp_j}\mu_{p'_j} x_{p\vp_j} x_{p'_j}}
{y_{p\vp_j} y_{p'_j}} ,
\nonumber\\
{\bf A}_L=\prod_{j=1}^L\frac{{\bf Z}_j}{y_{p\vp_j} y_{p'_j}},\quad
&&{\bf D}_L= \prod_{j=1}^L\frac {\mu_{p\vp_j}\mu_{p'_j}}{y_{p\vp_j} y_{p'_j}}.\quad 
\label{lcoeff}\ea
Consider the lattice with periodic boundary condition $\sigma_{L+1}=\sigma_1$, so that
\be
\sum_{i=1}^L n_i=\sum_{i=1}^L(\sigma_i-\sigma_{i+1})=\sigma_1-\sigma_{L+1}=0.
\ee
Then we have
$\prod_{j=1}^L{\bf Z}_j|\{n_i\}\rangle=|\{n_i\}\rangle$. Consequently,
if $\mu_{p_j}\mu_{p'_j}=1$, for all $j$, then $\Btau2$ has degenerate eigenspaces.
From (\ref{xyt}), we find the condition $\mu_{p}\mu_{p'}=1$ is satisfied if and only if
$(x_{p'},y_{p'},\mu_{p'})=(y\vp_p,\omega^j x\vp_p,\mu_p^{-1})$. This is
indeed the case for the superintegrable case with $j=N$.
\subsection{CSOS models}
In the context of the eight-vertex model Baxter \cite{BSOS,ABF} has introduced
the restricted solid-on-solid (rSOS) model , in which an interface is described
by assigning integer heights to the sites of a two-dimensional lattice,
while restricting the heights (or height differences) to a finite range.
Pearce and Seaton \cite{PK} chose a different restriction, choosing the
heights from some $\mathbb{Z}_N$ using the cyclic condition $N+1\equiv1$,
calling their model a cyclic solid-on-solid (CSOS) model. Here we shall
introduce other examples of CSOS models.

As mentioned above, it has been shown in \cite{BBP} that for
$(x_{q'},y_{q'},\mu_{q'})=(y\vp_q,\omega^\ell x\vp_q,\mu_q^{-1})$ and 
$(x_{p'},y_{p'},\mu_{p'})=(y\vp_p,\omega^j x\vp_p,\mu_p^{-1})$, the square
$U_{pp'qq'}(a,b,c,d)=0$ for $0\le a-d \le \ell-1$ and $\ell\le b-c \le N-1$; and 
also for $0\le d-c \le j-1$ and $j\le a-b \le N-1$. The diagonal block
$U^{(\ell,j)}_{pq}(a,b,c,d)$ describes a special case of the CSOS model
with $0\le a-d,b-c \le \ell-1$ and $0\le d-c,a-b \le j-1$, and depends
on $t_q$ and $t_p$ only. Its weights, which are left
implicit in \cite{BBP} are given in (\ref{squaresos}) in the Appendix.
The corresponding transfer matrices, denoted by $\btau_{\ell,j}$,
are special cases of the $\btau_{\ell}$ model acting on
restricted spaces with $0\le n_i\le j-1$.
\subsection{Outline of the paper}
In section 2, we consider the special case of such a CSOS model
with $\ell=2$, so that its monodromy matrix satisfies the Yang--Baxter
equation (\ref{YBEuUU}). The eigenvectors of this model are given
in section 2.2. In section 3, we use
the method of Baxter \cite{BaxterSB} to derive the Drinfeld polynomial
of the highest-weight representation which shows that the CSOS model
has $2^{m_E}$-fold degeneracy for some integers $m_E$, which will be
given later. This in turn means the existence of quantum loop algebras.
The generators of the quantum loop algebra given in (\ref{generators})
are the same as those given in \cite{APsu4}.  In section 4, we shall
present the proof of the Serre relations for these generators for the
CSOS models, which includes the superintegrable case.

Included in the Appendix A is rederivation of the decomposition of the
square of weights, as the notations used in \cite{BBP} are not conventional.
The corresponding functional relations for the product of two transfer matrices
given in \cite{BBP, BaxterSB} for these CSOS models are included here
in Appendix B. As the functional relations between the $\btau_j$-matrices
are direct consequences of fusion, it is shown in Appendix B.2, that
the T-system functional relations studied by many authors
\cite{KlPearce,JKlSuzuki,KNS} also hold for any $\btau_j$ model.
In Appendix C, the relationships between the coefficients of the monodromy
matrix (\ref{ABCD}), which only depend on the specific form of
the asymmetric 6-vertex model $R$-matrix, are given. Using these
relations, we also show in section 4, that
${\bf C}_{L-1}$, ${\bf B}_{L}$, ${\bf C}_{0}$ and ${\bf B}_{1}$
of the CSOS models are related to a $j^L$-dimensional representation
of  $U_{\rm q}(\widehat{\mathfrak{sl}}_2)$.
\section{CSOS models for \boldmath{$\ell=2$}}
\setcounter{equation}{\value{storeeqn}}
Using alternating horizontal and vertical rapidities,
\be\fl
(x_{q'},y_{q'},\mu_{q'})=(y\vp_q,\omega^2 x\vp_q,\mu_q^{-1})\quad \hbox{and}\quad
(x_{p'},y_{p'},\mu_{p'})=(y\vp_p,\omega^j x\vp_p,\mu_p^{-1}),
\label{qq'pp'}\ee
we have the decomposition of the square
\be
U_{pp'qq'}(a,b,c,d)\to U^{(2)}_{pp'q}(a,b,c,d){(t_q)}\to U^{(2,j)}_{pq}(a,b,c,d){(t_q/t_p)},
\ee
and from (\ref{u2})\footnote{We have also dropped the factors
$\mu_p/y_p$ in ${\bf U}^{(2,j)}_{pq}(a,b,b-1,a)$ and ${\bf U}^{(2,j)}_{pq}(a,b,b,a-1)$,
as they always appear in pairs in the transfer matrices and cancel out upon
multiplication.}, we find the nonvanishing elements to be
\ba {\bf  U}^{(2,j)}_{pq}(a,b,b,a) =
1-\omega^{1-j+a-b} t,&&\quad 0\le a-b\le j-1;\nonumber\\
{\bf U}^{(2,j)}_{pq}(a,b,b-1,a) =-\omega^{1-j}t(1-\omega^{1+a-b}),
&&\quad 0\le a-b\le j-2;\nonumber\\
{\bf  U}^{(2,j)}_{pq}(a,b,b,a-1) = 
1-\omega^{a-b-j},
&&\quad 1\le a-b\le j-1;\nonumber\\
{\bf  U}^{(2,j)}_{pq}(a,b,b-1,a-1) =  \omega^{1-j}( \omega^{a-b}- t),
&&\quad 0\le a-b\le j-1.
\label{csos2j}\ea
where we set $t=t_q /t_p$, so that the high-genus rapidities are replaced
by the usual rapidities with difference property.  The resulting transfer matrix is
\be\fl
[\btau_{2,j}(t_q/t_p)]_{\sigma,\sigma'}=
{\rm tr}[{\boldsymbol {\cal U}}^{(2,j)}(t)],\quad
{\boldsymbol {\cal U}}^{(2,j)}(t)\equiv\prod_{j=1}^L
 {\bf U}_{p,q}^{(2,j)}(\sigma\vp_j,\sigma\vp_{j+1},\sigma'_{j+1},\sigma'_{j}).
 \label{U2j}\ee
Since $\ell=2$, the Yang--Baxter equations (\ref{ybeUU}) or (\ref{YBEuUU})
hold for the monodromy matrix ${\boldsymbol {\cal U}}^{(2,j)}(t)$. As can be
seen from (\ref{csos2j}), the weights are simpler than those studied by
Pearce and Seaton and others \cite{PK,JMO}.
\subsection{Commutation relation for $\ell=2$.}
For $0\le d-c,a-b\le j-1$, using (\ref{csos2j}), we find the leading coefficients to be
\be\fl
{\bf A}_0=1,\quad
{\bf D}_0=\omega^{(1-j)L }\prod_{i=1}^L{\bf Z}^{(j)}_i,\quad
{\bf A}_L=\omega^{-jL }\prod_{i=1}^L{\bf Z}_i^{(j)},\quad
{\bf D}_L=\omega^{-jL },
\label{2coeff}\ee
where ${\bf Z}^{(j)}$ is the $j\times j$ diagonal matrix with elements
\be
{\bf Z}^{(j)}_{k,l}=\delta_{k,l}\omega^{k},\quad 0\le k,l\le j-1.
\label{Zj}\ee
From (\ref{csos2j}), we find
that the weights depend only on the difference of neighboring spins. As
the transfer matrices of the CSOS models commute with the spin shift operator 
 $\mbox{\mycal X}={\bf X}_1{\bf X}_2\cdots {\bf X}_L$, their eigenspaces
 split into $N$ blocks. In the block corresponding to the
eigenvalue $\omega^Q$ of the shift operator $\mbox{\mycal X}$,
the transfer matrix becomes ${\bf A}(t) +\omega^{-Q}{\bf D}(t)$.
Assuming cyclic boundary conditions and $L$ a multiple of $N$,
$L=p N$ for some integer $p$,
we find from (\ref{2coeff}) that the same commutation relations
\ba
&&[{\bf A}(x)+\omega^{-Q}{\bf D}(x),
{\bf C}_{0}^{(nN+Q)}{\bf B}_1^{(mN+Q)}]\,|\{n_j\}\rangle=0,
\nonumber\\
&&[{\bf A}(x)+\;\omega^{Q}\,{\bf D}(x),
{\bf B}_1^{(mN+Q)}{\bf C}_0^{(nN+Q)}]\,|\{n_j\}\rangle=0,
\nonumber\\
&&[{\bf A}(x)+\omega^{-Q}{\bf D}(x),
{\bf B}_L^{(mN+Q)}{\bf C}_{L-1}^{(nN+Q)}]\,|\{n_j\}\rangle=0,
\nonumber\\
&&[{\bf A}(x)+\;\omega^{Q}\,{\bf D}(x),
{\bf C}_{L-1}^{(nN+Q)}{\bf B}_L^{(mN+Q)}]\,|\{n_j\}\rangle=0,
\label{comm}\ea
hold as those given in (IV:49) and (IV:50) of \cite{APsu4}.\footnote{All
equations in \cite{APsu1},  \cite{APsu2}, or \cite{APsu4} are denoted here by prefacing
I, II, or IV to their equation numbers, all equations in~\cite{BBP} are denoted
by adding `BBP:' to their equation numbers, and all equations
in~\cite{BaxterSB} are denoted by adding `Baxter:' to their equation numbers.}  

Thus the generators of $L({\widehat{\mathfrak{sl}}}_2)$ for the ground-state sectors
in superintegrable models, as given in \cite{NiDe1, NiDe2, APsu1} for $Q=0$
and in \cite{APsu4} for $Q\ne0$, should also be generators for the CSOS model. 
To show that CSOS models with weights given by (\ref{csos2j}) support quantum
loop algebra $L({\widehat{\mathfrak{sl}}}_2)$ in all sectors, we must
prove that the generators satisfy the necessary Serre relations; this proof
will be given in section 4. We shall first present vectors, upon which these generators
generate eigenspaces spanned by $2^{m_E }$ eigenvectors having the same eigenvalue.
\subsection{Eigenvectors}
It is easily verified that Yang--Baxter equation (\ref{YBEuUU}) also holds
for the monodromy matrix ${\boldsymbol {\cal U}}(t_q)$ with different
vertical rapidities, as defined in (\ref{gtau2}).
Therefore, the well-known identities derived in \cite{STF, NiDe2} also hold for
this monodromy matrix, i.e.
\ba
 {\bf A}(x_0)\Bigg(\prod_{i=1}^R {\bf B}(x_i)\Bigg)=
\omega^R \Bigg[\Bigg(\prod_{i=1}^R f_{i0}\Bigg)
\Bigg(\prod_{i=1}^R {\bf B}(x_i)\Bigg) {\bf A}(x_0)\cr\hspace{80pt}
+\sum_{i=1}^R\Bigg(\prod_{k=1,k\ne i}^R f_{ki}\Bigg)g_{0i}
\Bigg(\prod_{k=0,k\ne i}^R {\bf B}(x_k)\Bigg) {\bf A}(x_i)\Bigg],\cr
{\bf D}(x_0)\Bigg(\prod_{i=1}^R {\bf B}(x_i)\Bigg)=
\omega^R \Bigg[\Bigg(\prod_{i=1}^R f_{0i}\Bigg)
\Bigg(\prod_{i=1}^R {\bf B}(x_i)\Bigg) {\bf D}(x_0)\cr\hspace{80pt}
-\sum_{i=1}^R\Bigg(\prod_{k=1,k\ne i}^R f_{ik}\Bigg)g_{0i}
\Bigg(\prod_{k=0,k\ne i}^R {\bf B}(x_k)\Bigg) {\bf D}(x_i)\Bigg],
\label{ADprodB}\end{eqnarray}
where we used the short-hand notations of \cite{NiDe2},
\be
\fl
f_{ik}=f\bigg(\frac{x_i}{x_k}\bigg),\quad
g_{ik}=g\bigg(\frac{x_i}{x_k}\bigg),\quad
f(z)=\frac{z-\omega}{\omega(z-1)},\quad
g(z)=\frac{1-\omega}{\omega(z-1)}.
\label{fg}\ee
Similarly, we also have
\ba
 {\bf A}(x_0)\Bigg(\prod_{i=1}^R {\bf C}(x_i)\Bigg)=
 \Bigg[\Bigg(\prod_{i=1}^R f_{0i}\Bigg)
\Bigg(\prod_{i=1}^R {\bf C}(x_i)\Bigg) {\bf A}(x_0)\cr\hspace{80pt}
+\sum_{i=1}^R\Bigg(\prod_{k=1,k\ne i}^R f_{ik}\Bigg)g_{i0}
\Bigg(\prod_{k=0,k\ne i}^R {\bf C}(x_k)\Bigg) {\bf A}(x_i)\Bigg],\cr
{\bf D}(x_0)\Bigg(\prod_{i=1}^R {\bf C}(x_i)\Bigg)=
 \Bigg[\Bigg(\prod_{i=1}^R f_{i0}\Bigg)
\Bigg(\prod_{i=1}^R {\bf C}(x_i)\Bigg) {\bf D}(x_0)\cr\hspace{80pt}
-\sum_{i=1}^R\Bigg(\prod_{k=1,k\ne i}^R f_{ki}\Bigg)g_{i0}
\Bigg(\prod_{k=0,k\ne i}^R {\bf C}(x_k)\Bigg) {\bf D}(x_i)\Bigg].
\label{ADprodC}\end{eqnarray}
In these equations, the subscripts are different from those of Nishino and
Deguchi \cite{NiDe2}, because of the difference in the $R$-matrices.

Consider the vector $|{\bf R}\rangle$ given by
\be 
|{\bf R}\rangle={\bf B}^{(\ell N-R)}_L\prod_{i=1}^{R}{\bf B}(x_i)|\Omega\rangle,
\quad |\Omega\rangle=|\{0\}\rangle=
\overset{L}{\underset{j=1}{\mbox{\Large$\otimes$}}}|0\rangle.
\label{vector1}\ee
Here $|\Omega\rangle$ is the state $|\{n_i\}\rangle$ with all
$n_i=\sigma_i-\sigma_{i+1}$ having the minimal value 0.
Then, using the commutation relations (\ref{commABn}), we find
\ba\fl
[{\bf A}(t)+\omega^{-Q}{\bf D}(t)|{\bf R}\rangle
={\bf B}^{(\ell N-R)}_L[{\bf A}(t)+\omega^{-Q-R}{\bf D}(t)]
\prod_{i=1}^R{\bf B}(x_i)|\Omega\rangle
\nonumber\\
+{(\omega-1){\bf B}^{(\ell N-R-1)}_L{\bf B}(t)\prod_{i=1}^R{\bf B}(x_i)
(\omega^{R}{\bf A}_L
-\omega^{-Q}{\bf D}_L)|\Omega\rangle}.
\label{ADR0}\ea
Note that the second term in (\ref{ADR0}) vanishes if either $R=\ell N$ or $R=nN-Q$.
If $R=\ell N$, the first line reproduces (\ref{ADR0}); if  $R=nN-Q$,  we can
use (\ref{2coeff}) and $\prod_{i=1}^L{\bf Z}_i^{(j)}|\Omega\rangle=|\Omega\rangle$,
which follows from (\ref{Zj}).
Also we find from (\ref{csos2j}) that
\ba 
{\bf A}(t)|\Omega\rangle=a(t)|\Omega\rangle,\quad a(t)=(1-\omega^{1-j}t)^L,
\nonumber\\
{\bf D}(t)|\Omega\rangle=d(t)|\Omega\rangle,
\quad d(t)=\omega^{(1-j)L}(1-t)^L.
\label{ADomega}\ea
Next set $x_0=t$ and define
\be
F(t)=\prod_{i=1}^{R}(t-\omega x_i)\label{defF},
\ee
so that, with $f_{ij}$ defined by (\ref{fg}),
\be \fl
\prod_{i=1}^R f_{0i}=\frac{F(t)}{F(\omega t)},\quad 
\prod_{i=1}^R f_{i0}=\omega^{-R}\frac{F(\omega^2t)}{F(\omega t)}, \quad 
\prod_{k=1,k\ne i}^R \frac{f_{ik}}{f_{ki}}=
\prod_{k=1,k\ne i}^R \frac{x_i-\omega x_k}{\omega x_i- x_k}.
\label{fi0}\ee
Then, using the identities (\ref{ADprodB}) and (\ref{ADomega}) in (\ref{ADR0}),
we obtain
\ba\fl
[{\bf A}(t)+\omega^{-Q}{\bf D}(t)]|{\bf R}\rangle
=\bigg[a(t)\frac{F(\omega^2t)}{F(\omega t)}
+\omega^{-Q}d(t)\frac{F(t)}{F(\omega t)}\bigg]|{\bf{R}}\rangle
\nonumber\\
+{\bf B}^{(\ell N-R)}_L
\sum_{i=1}^{R} g_{0i}
\prod_{k=0,k\ne i}^{R} {\bf B}(x_k)\prod_{k=1,k\ne i}^{R}[\omega(x_i-x_k)]^{-1}
\nonumber\\
\times{\bigg[a(x_i) \omega^{R}\prod_{k=1,k\ne i}^{R} (\omega x_i-x_k)
-d(x_i)\omega^{-Q}\prod_{k=1,k\ne i}^{R}( x_i-\omega x_k)\bigg]}|\Omega\rangle,
\label{ADR}\ea
where we have also used (\ref{fi0}) and (\ref{fg}).
If we choose the $x_i$ for $i=1,\cdots,R$ such that 
\be 
a(x_i)\omega^{R}\prod_{k=1,k\ne i}^{R}(\omega x_i- x_k)=
d(x_i)\omega^{-Q}\prod_{k=1,k\ne i}^{R} ( x_i-\omega x_k),
\ee
which are actually the Bethe Ansatz equations, then the second term
in (\ref{ADR}) vanishes. Then $|{\bf R}\rangle$ is an eigenvector of 
$[{\bf A}(t)+\omega^{-Q}{\bf D}(t)]$ with eigenvalue
\ba\fl
T_{\mathrm{CSOS}}=\tau_{2,j}(t)&=\bigg[a(t)\frac{F(\omega^2t)}{F(\omega t)}
+\omega^{-Q}d(t)\frac{F(t)}{F(\omega t)}\bigg]
\nonumber\\
&=\bigg[(1-\omega^{1-j}t)^L\frac{F(\omega^2t)}{F(\omega t)}
+\omega^{-Q+(1-j)L}(1-t)^L\frac{F(t)}{F(\omega t)}\bigg].
\label{ffcsos}\ea

Similarly, let
\be
|{\bar \Omega}\rangle=|\{j-1\}\rangle=
\overset{L}{\underset{j=1}{\mbox{\Large$\otimes$}}}|j-1\rangle
\ee
be the state $|\{n_i\}\rangle$ with all
$n_i=\sigma_i-\sigma_{i+1}$ having the maximal value $j-1$.
It is easy to see from (\ref{csos2j}) that
\ba 
{\bf A}(t)|{\bar \Omega}\rangle={\hat a}(t)|{ \bar \Omega}\rangle,\quad {\hat a}(t)=(1-t)^L,\nonumber\\
{\bf D}(t)|{\bar \Omega}\rangle={\hat d}(t)|{ \bar \Omega}\rangle,
\quad {\hat d}(t)=(1-\omega^{1-j}t)^L.
\ea
Consider now the vector
\be
|{\bf {\bar R}}\rangle={\bf C}^{(\ell N-R)}_{L-1}\prod_{i=1}^R{\bf C}(x_i)|{\bar\Omega}\rangle.
\label{vector2}\ee
Using the commutation relations
\ba
{\bf C}(x){\bf A}_L=\omega{\bf A}_L{\bf C}(x),\qquad
{\bf C}(x){\bf D}_L={\bf D}_L{\bf C}(x),
\nonumber\\
{\bf A}(x){\bf C}^{(n)}_{L-1}={\bf C}^{(n)}_{L-1}{\bf A}(x)+
(\omega-1)\omega^{1-n}x{\bf C}^{{(n-1)}}_{L-1}{\bf C}(x){\bf A}_L,
\nonumber\\
{\bf D}(x){\bf C}^{(n)}_{L-1}=\omega^{-n}{\bf C}^{(n)}_{L-1}
{\bf D}(x)-(\omega-1)\omega^{1-n}x{\bf C}^{(n-1)}_{L-1}{\bf C}(x){\bf D}_L,
\label{ADCn}\ea
we can easily show that,
if $R=\ell N$ or $R=n N+Q$ and $\{ x_1,x_2,\cdots,x_R\}$
satisfy the Bethe Ansatz equations
\be 
{\hat a}(x_i)\prod_{k=1,k\ne i}^{R}(x_i- \omega x_k)=\omega^{R-Q}
{\hat d}(x_i)\prod_{k=1,k\ne i}^{R} ( \omega x_i- x_k),
\label{bethe2}\ee
then $|{\bf \bar R}\rangle$ is also an eigenvector of $[{\bf A}(t)+\omega^{-Q}{\bf D}(t)]$
and has eigenvalue
\ba\fl
{T}_{\mathrm{CSOS}}=\tau_{2,j}(t)=&\bigg[{\hat a}(t)\frac{F(t)}{F(\omega t)}
+\omega^{-Q}{\hat d}(t)\frac{F(\omega^2t)}{F(\omega t)}\bigg]
\nonumber\\
&=\bigg[(1-t)^L\frac{F(t)}{F(\omega t)}
+\omega^{(1-j)L-Q}(1-\omega^{1-j}t)^L\frac{F(\omega^2t)}{F(\omega t)}\bigg],
\label{hfcsos}
\ea
where we have added $\omega^{(1-j)L}=1$ in the second line to make
the result also valid if $L\ne pN$. However, for $L\ne pN$, $|{\bar \Omega}\rangle$
does not satisfy the cyclic boundary condition, so that the vectors (\ref{vector2})
are not eigenvectors under that condition.

However, the vectors given by
\be 
|\hat{\bf R}\rangle={\bf B}^{(\ell N-R)}_1\prod_{i=1}^{R}{\bf B}(x_i)|\Omega\rangle,\quad
\omega^R=\omega^{Q-(1-j)L},\quad R\ne \ell N,
\label{vector3}\ee
with the $x_i$ for $i=1,\cdots,R$ satisfying the Bethe Ansatz equations
\be 
(1-x_i)^L\prod_{k=1,k\ne i}^{R}(x_i- \omega x_k)=
(1-\omega^{1-j}x_i)^L\prod_{k=1,k\ne i}^{R} ( \omega x_i- x_k),
\label{bethe3}\ee
can be shown to be eigenvectors with eigenvalues given by (\ref{hfcsos}).

From finite-size calculations, we find that these are not the only possibilities.
We must also introduce
\be
|\tilde{\bf R}\rangle={\bf B}^{(\ell N-R-n)}_L{\bf B}^{(n)}_1
\prod_{i=1}^{R}{\bf B}(x_i)|\Omega\rangle,\quad
\omega^R=\omega^{-Q-n}=\omega^{(j-1)L},
\label{vector4}\ee
with the $x_i$ satisfying the Bethe Ansatz equations
\be 
\omega^Q(1-x_i)^L\prod_{k=1,k\ne i}^{R}(x_i- \omega x_k)=
(1-\omega^{1-j}x_i)^L\prod_{k=1,k\ne i}^{R} ( \omega x_i- x_k).
\label{bethe4}\ee
Their eigenvalues are also given by (\ref{hfcsos}). 
 
We can summarize the results rewriting (\ref{ffcsos}) and (\ref{hfcsos}) as 
\be 
\tau_{2,j}(t)F(\omega t)=\omega^{-P_a}(1-t)^L{F(t)}
+\omega^{P_b}(1-\omega^{1-j}t)^L{F(\omega^2t)},
\label{tau2}\ee
where we must choose $-P_a=(1-j)L-Q \;\mod\;N$ and $P_b=0$ in (\ref{ffcsos})
and $P_a=0$ and $P_b=(1-j)L-Q \;\mod\;N$ in (\ref{hfcsos}).
Then the Bethe Ansatz equations become
\be\fl
(1-x_i)^L\prod_{k=1,k\ne i}^{R}(x_i- \omega x_k)=\omega^{P_a+P_b+R}
(1-\omega^{1-j}x_i)^L\prod_{k=1,k\ne i}^{R} ( \omega x_i- x_k).
\label{bethe}
\ee
These results include the superintegrable case when $j=N$.

The eigenvalues (\ref{tau2})  are easily seen to be independent of the
$\ell$ in (\ref{vector1}), (\ref{vector2}), (\ref{vector3}), and (\ref{vector4}),
which also shows the degeneracy of their eigenspaces. The smallest allowed
values of $\ell$ lead to the possible highest-weight vectors.

Thus, we have shown that the eigenvectors are degenerate, but we have not
yet demonstrated the Ising-like behavior with $2^{m_E}$-fold degeneracies.
To understand the degeneracy, we must calculate the highest-weight polynomials,
or the so-called Drinfeld polynomials \cite{Degu1,Degu2, NiDe2}. We shall use
the method of Baxter in \cite{BaxterSB} to determine these polynomials. As a
byproduct, the eigenvalues of all our CSOS models for any $\ell$ are explicitly given.
\section{Functional relations in CSOS models}
\setcounter{equation}{\value{storeeqn}} 
\subsection{Explicit formula for $\tau_{\ell,j}(t)$}
Using $-P_a+P_b=(1-j)L-Q$, we rewrite the functional relation (\ref{funlj}) as
\ba\fl
\tau_{2,j}(\omega^{\ell-1}t)\tau_{\ell,j}(t)
-\omega^{-P_a+P_b}(1-\omega^{\ell-1}t)^L(1-\omega^{\ell-1-j}t)^L\tau_{\ell-1,j}(t)
=\tau_{\ell+1,j}(t).
\label{funljp}\ea
We shall show by induction that
\ba\fl
\tau_{\ell,j}(t)=\sum_{n=0}^{\ell-1}\zeta_n^{\ell,j}(t),\qquad 2\le\ell\le N,
\label{tauljt}\\ \fl
\zeta_n^{\ell,j}(t)\equiv\frac{\omega^{nP_b-(\ell-1-n)P_a}F(t)F(\omega^{\ell} t)}
{F(\omega^{\ell-n-1} t)F(\omega^{\ell-n} t)}
\prod_{m=0}^{\ell-2-n}(1-\omega^m t)^L
\prod_{m=\ell-n}^{\ell-1}(1-\omega^{m-j} t)^L,
\label{tauljt2}\ea
for the eigenvalues $\tau_{\ell,j}(t)$ of $\btau_{\ell,j}(t)$.
It is easy to see that for $\ell=2$, we have
\ba\zeta_0^{2,j}(t)=\frac{\omega^{-P_a}F(t)}{F(\omega t)}(1- t)^L,
\quad \zeta_1^{2,j}(t)=\frac{\omega^{P_b}F(\omega^{2}t)}{F(\omega t)}(1- \omega^{1-j}t)^L,
\ea
so that $\tau_{2,j}$ is identical to (\ref{tau2}). Now assume (\ref{tauljt}) holds
for $\ell$ or smaller. Using (\ref{tauljt}) and (\ref{tauljt2}), we can easily show
\ba\fl
\zeta_0^{2,j}(\omega^{\ell-1}t)\zeta_0^{\ell,j}(t)=\omega^{-\ell P_a}
\frac{F(t)}{F(\omega^{\ell} t)}
\prod_{m=0}^{\ell-1}(1-\omega^m t)^L,
\label{taul1ja}\\
\fl \zeta_0^{2,j}(\omega^{\ell-1}t)\,\big[\tau_{\ell,j}(t)-\zeta_0^{\ell,j}(t)\big]
=\zeta_0^{2,j}(\omega^{\ell-1}t)\sum_{n=0}^{\ell-2}\zeta_{n+1}^{\ell,j}(t)
\nonumber\\
=\omega^{-P_a+P_b}(1-\omega^{\ell-1}t)^L(1-\omega^{\ell-1-j}t)^L\tau_{\ell-1,j}(t),
\label{taul1jb}\\
\fl \zeta_1^{2,j}(\omega^{\ell-1}t)\tau_{\ell,j}(t)=
\sum_{n=0}^{\ell-1}\frac{\omega^{(n+1)P_b-(\ell-n-1)P_a}F( t)F(\omega^{\ell+1} t)}
{F(\omega^{\ell-n-1} t)F(\omega^{\ell-n} t)}
\nonumber\\
\times\prod_{m=0}^{\ell-2-n}(1-\omega^m t)^L
\prod_{m=\ell-n}^{\ell}(1-\omega^{m-j} t)^L.
\label{taul1j}\ea
Also, $\tau_{2,j}(\omega^{\ell-1}t)\tau_{\ell,j}(t)$ in (\ref{funljp}) is the sum
of the left-hand sides of (\ref{taul1ja}), (\ref{taul1jb}) and  (\ref{taul1j}), while
the right-hand side of (\ref{taul1jb}) cancels the second term of (\ref{funljp}).
Therefore, replacing $n=n'-1$ in  (\ref{taul1j}),we obtain the desired result
\ba\fl
\tau_{\ell+1,j}(t)=\sum_{n=0}^{\ell}
\frac{\omega^{nP_b-(\ell-n)P_a}F(t)F(\omega^{\ell+1} t)}
{F(\omega^{\ell-n} t)F(\omega^{\ell-n+1} t)}
\prod_{m=0}^{\ell-1-n}(1-\omega^m t)^L
\prod_{m=\ell+1-n}^{\ell}(1-\omega^{m-j} t)^L.
\ea
This proves (\ref{tauljt}) holds for all $\ell\le N$.
\subsection{Functional relations for the transfer matrices}
Following the method of Baxter in chapter 6 of \cite{BaxterSB}, we introduce
\ba\fl
T_q=N^{\halfs L} \frac{(1-x_q/y_{p})^L}
{(1-x^N_q/y^N_{p})^L}{\cal T}(x_q,y_q),\quad
{\hat T}_{q'}=N^{\halfs L}\frac{(1-\omega^{-j}y_q/x_{p})^L}
{(1-y_q^N/x_{p}^N)^L}{\hat{\cal T}}(y_q,\omega^\ell x_q).
\label{calT}\ea
It has been shown by Baxter \cite{BaxterSB} or can be seen from (\ref{Apq}) that ${\cal T}(x_q,y_q)$ and 
${\hat{\cal T}}(y_q\omega^\ell,x_q)$ are polynomials in $x_q$ and $y_q$.
Let $t=t_q/t_p$, so that (\ref{funttsos}) for $\ell=1,\cdots,N$ becomes
\ba\fl
{\cal T}(x_q,y_q){\hat{\cal T}}(y_q,\omega^\ell x_q)=
\Bigg(\prod_{m=\ell}^{N-1}(1-\omega^{m-j} t)^L\Bigg)\btau_{\ell,j}(t)\nonumber\\
+\omega^{\ell (P_b-P_a)}\Bigg(\prod_{m=0}^{\ell-1}(1-\omega^{m} t)^L\Bigg)
\btau_{N-\ell,j}(\omega^\ell t).
\label{fun2}\ea
Now substituting (\ref{tauljt}) into (\ref{fun2}), we find that its eigenvalues become
\ba\fl
{\cal T}(x_q,y_q){\hat{\cal T}}(y_q,\,\omega^\ell x_q)=\omega^{\ell P_b}
\Bigg(\prod_{m=0}^{N-1-j}(1-\omega^{m} t)^L\Bigg) F(t)F(\omega^\ell t)\, t^{P_a+P_b}\,{\cal P}(t^N),
\label{fun3}\ea
where 
\ba
t_q^{P_a+P_b}{\cal P}(t^N)=\omega^{-P_b}\sum_{k=0}^{N-1}\frac{\omega^{-k(P_a+P_b)}}
{F(\omega^k t)F(\omega^{k+1} t)}\prod_{n=1}^{j-1}(1-\omega^{k-n}t)^L.
\label{Bamp}\ea
For $j=N$, this reduces to the result for the superintegrable case examined by Baxter
in \cite{BaxterSB}. It is also easily seen that the degree of ${\cal P}(t^N)$ is
$m_E=\lfloor \big(L(j-1)-2R-P_a-P_b\big)/N\rfloor$.
From (\ref{xyt}) we find that we may write
\ba 
{\cal P}(t_q^N/t_p)=G(\lambda_q)G(\lambda^{-1}_q),\qquad \lambda_q=\mu_q^N.
\label{PGG}\ea
\subsection{Analysis of the transfer matrices and their eigenvalues}
Using (Baxter:2.22), we find
\ba
T(\omega x_q,\omega^{-1} y_q)=\left[\frac{(y_p-\omega x_q)(y_q-\omega^{1+j} x_p)}
{(y_q-\omega x_p)\omega(y_p- x_q)}\right]^L{\mbox{\mycal X}}^{-1} T( x_q,y_q),\nonumber\\
{\hat T}(\omega y_q,\omega^{-1} x_q)=\left[\frac{(x_q-\omega y_p)(\omega^{j} x_p-\omega y_q)}
{(x_p-y_q)\omega(x_q-\omega y_p)}\right]^L {\mbox{\mycal X}}^{-1}  {\hat T}(y_q, x_q).
\ea
Similarly, we can show 
\ba
T(\omega^{-1} x_q,\omega y_q)=\left[\frac{(y_q- x_p)(\omega y_p- x_q)}
{(y_p- x_q)(y_q-\omega^{j} x_p)}\right]^L{\mbox{\mycal X}} \,T( x_q,y_q),\nonumber\\
{\hat T}(\omega^{-1} y_q,\omega x_q)=\left[\frac
{(\omega x_p-y_q)(x_q- y_p)}
{(x_q- y_p)(\omega^{j} x_p- y_q)}\right]^L {\mbox{\mycal X}} \,{\hat T}(y_q, x_q).
\ea
Therefore, when the shift operator ${\mbox{\mycal X}}$ is replaced by
$\omega^{-Q}$, the rescaled polynomial transfer matrices defined in (\ref{calT})
restricted to the sector corresponding to $Q$, satisfy
\ba
{\cal T}(\omega x_q,\omega^{-1} y_q)=\omega^{P_a-P_b}&\left[\frac{1-\omega^{-1-j}y_q/x_p}
{1-\omega^{-1}y_q/x_p}\right]^L&{\cal T}( x_q,y_q),\label{shiftT1}\\
{\hat {\cal T}}(\omega y_q,\omega^{-1} x_q)=\omega^{P_a-P_b}&\left[\frac{1-\omega^{-j}y_q/x_p}
{1-y_q/x_p}\right]^L&{\hat{\cal T}}( y_q,x_q),\label{shifthT1}\\
{\cal T}(\omega^{-1} x_q,\omega y_q)=\omega^{-P_a+P_b}&\left[\frac{1-y_q/x_p}
{1-\omega^{-j}y_q/x_p}\right]^L&{\cal T}( x_q,y_q),\label{shiftT2}\\
{\hat {\cal T}}(\omega^{-1} y_q,\omega x_q)=\omega^{-P_a+P_b}&\left[\frac{1-\omega^{-1}y_q/x_p}
{1-\omega^{-1-j}y_q/x_p}\right]^L&{\hat{\cal T}}( y_q,x_q),
\label{shifthT2}\ea
where $P_a-P_b=(j-1)L+Q$, since we have chosen the multiplication from up-to-down. 
If one prefers Baxter's convention, one needs to make the change of $Q\to-Q$.
When $j=N$, this  reduces to the superintegrable case in (Baxter:6.5) with $r=0$ as it should.

For $(x_{p'},y_{p'},\mu_{p'})=(y_p,\omega^j x_p,\mu_p^{-1})$, we can use (BBP:2.20) and
(BBP:2.44) to show 
\ba\fl
\frac{f_{p'q}f_{pr}}{f_{pq}f_{p'r}}=\frac{(1-x_q/y_{p})(1-\omega^{-j}x_r/x_{p})(1-x^N_q/x^N_{p})
(1-x^N_r/y^N_{p})}
{(1-x^N_q/y^N_{p})(1-x^N_r/x^N_{p})(1-\omega^{-j}x_q/x_{p})(1-x_r/y_{p})}
\nonumber\\
\times\prod_{n=N-j}^{N-1}\frac{(1-\omega^{n}t_r/t_{p})(1-\omega^{n}x_q/x_{p})(1-\omega^{n}y_q/x_{p})}
{(1-\omega^{n}t_q/t_{p})(1-\omega^{n}x_r/x_{p})(1-\omega^{n}y_r/x_{p})},
\ea
with $f_{pq}$ defined in \cite[equation~(13)]{BPA}. The commutation relation
(Baxter:2.12) can then be rewritten for the rescaled transfer matrices in (\ref{calT}) as
\ba\fl
{\cal T}( x_q,y_q){\hat{\cal T}( x_r,y_r)}={\cal T}( x_r,y_r){\hat{\cal T}( x_q,y_q)}
\nonumber\\
\times\prod_{n=N-j}^{N-1}\Bigg[\frac{(1-\omega^{n}t_r/t_{p})(1-\omega^{n}x_q/x_{p})(1-\omega^{n}y_q/x_{p})}
{(1-\omega^{n}t_q/t_{p})(1-\omega^{n}x_r/x_{p})(1-\omega^{n}y_r/x_{p})}\Bigg]^L.
\ea
This relation holds for any $q$ and $r$, which suggest that
\ba
{\hat{\cal T}( x_q,y_q)}=\kappa\,{\cal T}( x_q,y_q)\prod_{n=N-j}^{N-1}
\Bigg[\frac{(1-\omega^{n}t_q/t_{p})}{(1-\omega^{n}x_q/x_{p})(1-\omega^{n}y_q/x_{p})}\Bigg]^L.
\label{hatTt}\ea
where $\kappa$ is some constant. Now we can use (\ref{shiftT1}) to (\ref{shifthT2}) and
(\ref{hatTt}) to find the transfer matrix eigenvalues, such that (\ref{fun3}) is satisfied.
Let us write 
\ba
{\cal T}( x_q,y_q)=x_q^{P_a}y_q^{P_b}\lambda_q^{-P_c}G(\lambda_q^{-1})F(t_q)
\prod_{n=0}^{N-j-1}\frac{(1-\omega^{n}t_q/t_{p})^L}{(1-\omega^{n}y_q/x_{p})^L},
\label{calTxy}\ea
where $P_a$, $P_b$ and $P_c$ are integers in the interval $0\le P_a,P_b,P_c\le N-1$.
We suggest that (\ref{calTxy}) is still a polynomial as the zeroes in the denominators are
cancelled out by the zeroes in the numerator.
If $x_p=\omega^n  y_q$, then $\lambda^{-1}_p=\lambda_q$, so that
$\lambda_p=\lambda^{-1}_q$ or $y_p^N=x_q^N$. Thus we find $t_p^N=t_q^N$.
There is an $N$-sheet branch cut structure for variables $x_q,y_q$ and $t_q$,
but we may choose the sheet so that $t_p=\omega^n  t_q$.
Thus ${\cal T}( x_q,y_q)$ is free of poles. The expression (\ref{calTxy}) for
${\cal T}( x_q,y_q)$ can be easily shown to satisfy (\ref{shiftT1}) and (\ref{shiftT2}). 
From (\ref{calTxy}), we have 
\ba
{\cal T}( y_q,x_q)=y_q^{P_a}x_q^{P_b}\lambda_q^{P_c}G(\lambda_q)F(t_q)
\prod_{n=0}^{N-j-1}\frac{(1-\omega^{n}t_q/t_{p})^L}{(1-\omega^{n}x_q/x_{p})^L}.
\label{calTyx}\ea
Now we use (\ref{hatTt}) to obtain
\ba\fl
{\hat{\cal T}( y_q,x_q)}=\kappa y_q^{P_a}x_q^{P_b}\lambda_q^{P_c}G(\lambda_q)F(t_q)
\prod_{n=0}^{N-1}\frac{(1-\omega^{n}t_q/t_{p})^L}{(1-\omega^{n}x_q/x_{p})^L}
\prod_{n=N-j}^{N-1}\frac1{(1-\omega^{n}y_q/x_{p})^L}.
\label{hatTyx}\ea
Using (\ref{xyt}), we may write
\ba\fl
\prod_{n=0}^{N-1}\frac{1-\omega^{n}t_q/t_{p}}{1-\omega^{n}x_q/x_{p}}
=
\frac{1-t^N_q/t^N_{p}}{1-x^N_q/x^N_{p}}=-\frac{\mu^N_p x^N_p}{k'y_p^N}\bigg(1-\frac{ y^N_q}{x^N_{p}}\bigg)
=\kappa^{-1/L}\prod_{n=0}^{N-1}\bigg(1-\frac{\omega^n y_q}{x_{p}}\bigg)
\ea
with $\kappa^{1/L}=-k'y_p^N/\mu^N_p x^N_p$, so that (\ref{hatTyx}) becomes 
\ba
{\hat{\cal T}( y_q,x_q)}= y_q^{P_a}x_q^{P_b}\lambda_q^{P_c}G(\lambda_q)F(t_q)
\prod_{n=0}^{N-1-j}{(1-\omega^{n}y_q/x_{p})^L}.
\label{hatTyx2}\ea
It can again easily verified that ${\hat{\cal T}( y_q,x_q)}$ as given by (\ref{hatTyx2})
satisfies the relations (\ref{shifthT1}) and (\ref{shifthT2}).
Furthermore, substituting (\ref{calTxy}) and (\ref{hatTyx2}) into (\ref{fun3}), we find it becomes an identity.
As explained in \cite{BaxterSB}, we find from (\ref{xyt}) that, in the limit $\mu_q\to 0$,
$x_q\to \infty$, while $y_q$, $x_q\mu_q$ remain finite. This means that the weights
in (\ref{weights}) are finite, and so are $T_q$ and ${\hat T}_{q'}$.
From (\ref{calT}), we find then that ${\cal T}(x_q,y_q)$ diverges no faster
than $x_q^{(N-1)L}$ and ${\hat{\cal T}}(y_q,x_q)$ stays finite. In this limit,
we find from (\ref{hatTyx2}) and (\ref{calTxy}) that
\ba
{\hat{\cal T}}( y_q,x_q)\to x_q^{P_b-NP_c+R},
\quad
{\cal T}( x_q,y_q)\to x_q^{P_a+NP_c+N m_E+R+(N-j)L}.
\label{mu0}\ea
Thus if we choose the integer $P_c$ such that
\be P_b+R\le NP_c\le (j-1)L-m_EN-P_a-R,\label{Pc}\ee
then ${\hat{\cal T}}(y_q,x_q)$ is finite, 
and ${\cal T}(x_q,y_q)$ is $\mathrm{O}(x_q^{(N-1)L})$. 
Similarly, in the limit $\mu_q\to \infty$, we find from
(\ref{xyt}) that $y_q\to \infty$, while $x_q$, $y_q/\mu_q$ remain finite, such that
\ba
{\cal T}( x_q,y_q)\to y_q^{P_b-NP_c+R},
\quad
{\hat{\cal T}}( y_q,x_q)\to y_q^{P_a+NP_c+N m_E+R+(N-j)L}.
\label{muinf}\ea
The condition in (\ref{Pc}) then guarantees that ${\cal T}(x_q,y_q)$ is finite and
${\hat{\cal T}}(y_q,x_q)$ diverges no faster than $ y_q^{L(N-1)}$, as it should.

\section{Serre relations of the quantum loop algebra $L({\mathfrak{sl}}_2)$
for the generators}
\setcounter{equation}{\value{storeeqn}} 
The superintegrable chiral Potts models are found to have Ising-like 
spectra \cite{vonGehlen1985, AMP,Baxter1989}, and here we have shown that
our CSOS models behave similarly.

For $Q=0$ and $L$ a multiple of $N$, it has been shown
\cite{NiDe1,NiDe2} that the eigenspace in the superintegrable case supports a quantum
loop algebra $L({\mathfrak{sl}}_2)$. Furthermore, this loop algebra
can be decomposed into $r=\lfloor (N-1)L/N\rfloor$ simple ${\mathfrak{sl}}_2$ algebras
\cite{APsu1,APsu2,APsu4}.

For the $Q\ne 0$ cases, we have assumed in \cite{APsu4} that the Serre relations hold.
Even though, we have shown these relation to hold when operated on some special
vectors, see Appendix B of \cite{APsu4}, and also tested them extensively by
computer for small systems, a proof was still lacking. In this section,
we shall present the proof for the CSOS model, which includes the superintegrable
case as a special case.

We shall first show that ${\bf C}_{L-1}$, ${\bf B}_{L}$, ${\bf C}_{0}$ and ${\bf B}_{1}$
for the CSOS models are related to a $j^L$-dimensional representation of the affine
quantum group $U_{\rm q}(\widehat{\mathfrak{sl}}_2)$. Therefore, the higher-order
quantum Serre relation in Proposition 7.1.5 of Lusztig \cite{Lusztig} holds also
for the CSOS model. From (\ref{csos2j}) and  (\ref{ABCD}), we find
\begin{eqnarray}
\fl
{\bf B}_L=
\omega^{-jL}\sum_{n=1}^L\Bigg(\prod_{m=1}^{n-1}\bZ^{(j)}_m\Bigg)
\,\hat{\mbox{\bfrak f}}_n,\quad
{\bf C}_0=\sum_{n=1}^L\omega^{(n-1)(1-j)}
\Bigg(\prod_{m=1}^{n-1}\bZ^{(j)}_m\Bigg)\,
\hat{\mbox{\bfrak e}}_n,\nonumber\\
\fl
{\bf B}_1=
\omega^{-j}\sum_{n=1}^L\,\omega^{(L-n)(1-j)}\hat{\mbox{\bfrak f}}_n
\prod_{m=n+1}^{L}\bZ^{(j)}_m,\quad
{\bf C}_{L-1}=\omega^{-j(L-1)}\sum_{n=1}^L\,\hat{\mbox{\bfrak e}}_n
\prod_{m=n+1}^{L}\bZ^{(j)}_m,
\label{BC}\end{eqnarray}
where $\bZ^{(j)}$ is the $j\times j$ diagonal matrix obtained by deleting
the last $N-j$ columns and rows of $\bZ$.  Let $\bX^{(j)}$ denote the
$j\times j$ singular matrix obtained by deleting the last $N-j$ columns
and rows of $\bX$,\footnote{For the superintegrable case with $j=N$,
$\bZ$ is unchanged, but we choose $\bX^{(N)}$ to be singular, with
$\bX^{(N)}_{1,N}=0$, such that $\bQ\bX^{(N)}=\mathrm{q}\bX^{(N)}\bQ$,
$\bQ\equiv(\bZ^{(j)})^{1/2}$. Details like this, needed for $N$ even, are
missing in our early version of the proof of the Serre relations \cite{APserre}
for the superintegrable case.}
The other operators in (\ref{BC}) are
\ba
\hat{\mbox{\bfrak e}}_n=(\bX_n^{(j)})^T(1-\omega^{-j}\bZ^{(j)}_n),\quad
\hat{\mbox{\bfrak f}}_n=(1-\bZ^{(j)}_n)\bX^{(j)}_n.
\label{hehf}\ea
\subsection{Representations of $U_{\rm q}({\mathfrak{sl}}_2)$}
The equations in Appendix C are valid for any $L$. If we let $L=1$, the three equations
in (\ref{CB0L}) and (\ref{C0B1}) become one,
\ba
{\bf C}_0{\bf B}_1-\omega{\bf B}_1{\bf C}_0
=(1-\omega)[{\bf D}_1{\bf A}_0-{\bf D}_0{\bf A}_1].
\label{CBL1}\ea
From (\ref{BC}), we find
${\bf B}_1=\omega^{-j}\hat{\mbox{\bfrak f}}_1$, and ${\bf C}_0=\hat{\mbox{\bfrak e}}_1$.
From (\ref{2coeff}), we get
${\bf D}_1{\bf A}_{0}=\omega^{-j}$ and ${\bf D}_0{\bf A}_{1}=\omega^{1-2j}({\bf Z}_1^{(j)})^2$.
Next we let ${\rm q}^2=\omega$, ${\bf Z}_1^{(j)}=\bQ_1^{\;2}$ and
\ba\fl{\bf T}^{-2}=\omega^{1-j}({\bf Z}_1^{(j)})^2,\quad
{\bf B}_1=({\rm q}-{\rm q}^{-1})\mu{\bf T}^{-\halfs}{\bf F},\quad
 {\bf C}_0=({\rm q}-{\rm q}^{-1})\nu{\bf E}{\bf T}^{-\halfs}, 
 \label{TFE}\ea
where $\mu=-\omega^{-j}\rq^{\halfs(j-1)}$ and $\nu=\rq^{\halfs(-j-1)}$.
Then we have
\ba{\bf T}^{-1}{\bf F}={\rm q}^2{\bf F}{\bf T}^{-1},\quad 
{\bf T}{\bf E}={\rm q}^2{\bf E}{\bf T}. \label{comTFE}\ea
Substituting (\ref{TFE}) into (\ref{CBL1}), and using (\ref{comTFE}), we find
\ba
({\bf E}{\bf F}-{\bf F}{\bf E})=({\bf T}-{\bf T}^{-1})\big/ ({\rm q}-{\rm q}^{-1}).
\ea
Thus ${\bf T}$, ${\bf E}$ and ${\bf F}$ are generators of $U_{\rm q}({\mathfrak{sl}}_2)$.

\subsection{Representations of $U_{\rm q}({\widehat{\mathfrak{sl}}}_2)$}
For $L\ge2$, we find from (\ref{2coeff}) that
\ba{\bf D}_L{\bf A}_{0}=\omega^{-jL},\quad {\bf D}_0{\bf A}_{L}=\omega^{(1-2j)L}\prod_{i=1}^L({\bf Z}_i^{(j)})^2
=\omega^{-jL}{\bf T}_0^{-2},\ea
defining generator ${\bf T}_0$. We obtain further generators from
\ba
{\bf B}_L=({\rm q}-{\rm q}^{-1})\mu\vp_0{\bf T}_0^{-\halfs}{\bf F}\vp_0,\quad 
{\bf C}_0=({\rm q}-{\rm q}^{-1})\nu\vp_0{\bf E}\vp_0{\bf T}_0^{-\halfs},\quad
 {\bf T}\vp_1={\bf T}^{-1}_0,
\nonumber\\
{\bf B}_1=({\rm q}-{\rm q}^{-1})\mu\vp_1{\bf T}_0^{-\halfs}{\bf E}\vp_1,
\quad {\bf C}_{L-1}=({\rm q}-{\rm q}^{-1})\nu\vp_1{\bf F}\vp_1{\bf T}_0^{-\halfs},
\label{FE01}\ea
where 
\ba \mu_0=-\omega^{-jL}\rq^{-\halfs(j-1)},\quad
&\nu_0=\rq^{\halfs(j-1)-1},\nonumber\\
\mu_1=-\omega^{-j}\rq^{\halfs(j-1)-L(j-1)},\quad
&\nu_1=\omega^{-j(L-1)}\rq^{-\halfs(j+1)+L(j-1)}.
\label{munu01}\ea
It is easily shown that with $ \alpha_{ii}=2$ and $\alpha_{ij}=-2$,
we have for $0\le i,j\le 1$ the relations
\ba
{\bf T}_i^{-1}{\bf F}\vp_j={\rm q}^{\alpha_{ij}}{\bf F}\vp_j{\bf T}_i^{-1},\quad 
{\bf T}\vp_i{\bf E}\vp_j={\rm q}^{\alpha_{ij}}{\bf E}\vp_j{\bf T}\vp_i,
 \label{comFE01}\ea
Substituting (\ref{FE01}) into (\ref{CB0L}), and using (\ref{munu01}) and
(\ref{comFE01}), we find
\ba
{\bf E}_i{\bf F}_j-{\bf F}_j{\bf E}_i=
\delta_{i,j}({\bf T}\vp_i-{\bf T}_i^{-1})\big/({\rm q}-{\rm q}^{-1}).
\ea
We also need the following definitions for the scaled powers,
\ba
\fl
{\bf B}_\ell^{(n)}=\frac{{\bf B}_\ell^{\,n}}{[n]!},\quad
{\bf C}_\ell^{(n)}=\frac{{\bf C}_\ell^{\,n}}{[n]!},\quad
[n]!=\prod_{\ell=1}^n\frac{1-\omega^{\ell}}{1-\omega},
\nonumber\\
\fl
{\bf E}_i^{(n)}=\frac{{\bf E}_i^{\,n}}{[n]_\rq!},\quad
{\bf F}_i^{(n)}=\frac{{\bf F}_i^{\,n}}{[n]_\rq!},\quad
[n]_\rq!=\prod_{\ell=1}^n\frac{\rq^{\ell}-\rq^{-\ell}}{\rq-\rq^{-1}},\quad
[n]!={\rm q}^{\halfs n(n-1)}[n]_{\rm q}!,
\label{BEn}\ea
from which we find
\ba
{\bf C}\vp_0{\bf B}_1^{(3)}\propto{\rm q}^6{\bf E}\vp_0{\bf E}^{(3)}_1{\bf T}_0^{-2},
\quad
&&{\bf B}\vp_1{\bf C}\vp_0{\bf B}^{(2)}_1\propto
{\rm q}^6{\bf E}\vp_1{\bf E}\vp_0{\bf E}^{(2)}_1{\bf T}_0^{-2},\nonumber\\
{\bf B}^{(2)}_1{\bf C}\vp_0{\bf B}\vp_1\propto
{\rm q}^4{\bf E}\vp_0{\bf E}^{(3)}_1{\bf T}_0^{-2},
\quad
&&{\bf B}^{(3)}_1{\bf C}\vp_0\propto{\bf E}^{(3)}_1{\bf E}\vp_0{\bf T}_0^{-2}.
\ea
Similar relations hold for other combinations. 
Consequently, after canceling out the $\omega$-factors in (\ref{serremd1}) and (\ref{serremd2}), these relations become
\ba
{\bf E}\vp_{i}{\bf E}^{(3)}_{j}-{\bf E}\vp_{j}{\bf E}\vp_{i}{\bf E}^{(2)}_{j}
+{\bf E}^{(2)}_{j}{\bf E}\vp_{i}{\bf E}\vp_{j}-{\bf E}^{(3)}_{j}{\bf E}\vp_{i}=0,
\nonumber\\
{\bf F}\vp_{i}{\bf F}^{(3)}_{j}-{\bf F}\vp_{j}{\bf F}\vp_{i}{\bf F}^{(2)}_{j}
+{\bf F}^{(2)}_{j}{\bf F}\vp_{i}{\bf F}\vp_{j}-{\bf F}^{(3)}_{j}{\bf F}\vp_{i}=0,
\label{serre2}\ea
for $i\ne j$ and $i,j=0,1$.
This also shows that the $j^L\times j^L$ matrices ${\bf C}_0$, ${\bf B}_1$,
${\bf C}_{L-1}$ and ${\bf B}_L$ are related to the highest-weight
representations of the affine quantum group $U_{\rm q}({\widehat{\mathfrak{sl}}}_2)$,
leaving out the discussion of the coproduct and other operators here.
Consequently the higher-order quantum Serre relations in {\bf 7.1.6} of
Lusztig \cite{Lusztig} hold. If we define
\ba
(\rq-\rq^{-1}){\mbox{\bfrak e'}}=(\bX^{(j)})^T{(\rq^j\bQ^{-1}-\rq^{-j}\bQ)},\nonumber\\
(\rq-\rq^{-1}){\mbox{\bfrak f'}}={(\bQ-\bQ^{-1})}\bX^{(j)},\qquad
{\mbox{\bfrak t'}}=\rq^{j-1}\bQ^{-2},
\label{e'f'}\ea
then the generators in (\ref{FE01}) are explicitly given as
\ba\fl {\bf E}_0=\sum_{n=0}^L\rq^{-n(j-1)}\prod_{m=1}^{n-1}
 {\mbox{\bfrak t'}}_{m}^{-\halfs}{\mbox{\bfrak e'}}_{n}
\prod_{m=n+1}^{L} {\mbox{\bfrak t'}}_{m}^{\halfs},\quad
{\bf F}_0=\sum_{n=0}^L\rq^{n(j-1)}\prod_{m=1}^{n-1} 
{\mbox{\bfrak t'}}_{m}^{-\halfs}{\mbox{\bfrak f'}}_{n}
\prod_{m=n+1}^{L} {\mbox{\bfrak t'}}_{m}^{\halfs},\nonumber\\
\fl
 {\bf F}_1=\sum_{n=0}^L\rq^{-n(j-1)}\prod_{m=1}^{n-1} 
 {\mbox{\bfrak t'}}_{m}^{\halfs}{\mbox{\bfrak e'}}_{n}
\prod_{m=n+1}^{L} {\mbox{\bfrak t'}}_{m}^{-\halfs},\quad
{\bf E}_1=\sum_{n=0}^L\rq^{n(j-1)}\prod_{m=1}^{n-1}
 {\mbox{\bfrak t'}}_{m}^{\halfs}{\mbox{\bfrak f'}}_{n}
\prod_{m=n+1}^{L} {\mbox{\bfrak t'}}_{m}^{-\halfs}.
\ea
\subsection{Serre relation for the generators of the loop algebra}
As in \cite{APsu4}, the generators of the loop algebra are given by
\ba
{\bf x}_{0,Q}^+\propto{\bf C}_0^{(N+Q)}{\bf B}_{1}^{(Q)},\quad
{\bf x}_{1,Q}^-\propto{\bf C}_0^{(Q)}{\bf B}_{1}^{(N+Q)},\nonumber\\
{\bf x}_{-1,Q}^+\propto{\bf C}_{L-1}^{(N+Q)}{\bf B}_{L}^{(Q)},\quad
{\bf x}_{0,Q}^-\propto{\bf C}_{L-1}^{(Q)}{\bf B}_{L}^{(N+Q)}.
\label{generators}\ea
For $Q=0$, each term in the Serre relation is a product of 4 operators.
For $Q\ne0$, each term in the Serre relation is a product of 8 operators.
To prove the $Q\ne0$ case, we need to move factors around.
We shall first prove the identities
\be
 {\bf C}_0^{(N+Q)}{\bf B}_1^{(Q)}{\bf C}_0^{(Q)}=
 {\bf C}_0^{(Q)}{\bf B}_1^{(Q)}{\bf C}_0^{(N+Q)}.
\label{CBC1}\ee

As in Chapter 7 of Lusztig \cite{Lusztig}, but now for the cyclic case with $\rq^{2N}=1$,
we define
 \be\fl
f_{i,j,n,m;e}=f_{n,m;e}=\sum_{r+s=m} (-1)^r \rq^{er(2n-m+1)}
\theta_i^{(r)}\theta_j^{(n)}\theta_i^{(s)},\quad i,j=0,1,
\quad j\ne i,
\label{fij}\ee
where we may choose $e=\pm1$, $\theta_i={\bf E}_i$ or $\theta_i={\bf F}_i$, and
$\theta_i^{(r)}=\theta_i^{r}/[r]_\rq !$.
It is shown by Lusztig in Proposition 7.15.(b) \cite{Lusztig} that if $m>2n$, then $f_{n,m;e}=0$.
For $n=1$, and $m=3$, this is the usual quantum Serre relation given in (\ref{serre2}).
We follow the steps of Lusztig in his proof. 
Since $f_{n,m-\ell;e}=0$ for $\ell\le m-2n-1$, we have
\be
g=\sum_{\ell=0}^{m-2n-1}(-1)^\ell \rq^{\ell(1-m)} f\vp_{n,m-\ell;1}\theta_i^{(\ell)}=0.
\ee
Using (\ref{fij}), we find
\ba
g&=\sum_{\ell=0}^{m-2n-1}\sum_{r+s'=m-\ell}(-1)^{\ell+r} \rq^{\ell(1-m)+r(2n-m+\ell+1)}
 \theta_i^{(r)}\theta_j^{(n)}\theta_i^{(s')}\theta_i^{(\ell)}\nonumber\\
 &=\sum_{s=0}^m c_{s}\,
 \theta_i^{(m-s)}\theta_j^{(n)}\theta_i^{(s)}=0,\quad s=s'+\ell,\quad r=m-s,
 \label{gh}\ea
where
\be
 \fl\qquad c_{s}=\sum_{\ell=0}^{m-2n-1}(-1)^{\ell+m-s} \rq^{\ell(1-s)+(m-s)(2n-m+1)}
 \sfactor s{\ell}_\rq,\quad \sfactor s{\ell}_\rq=\frac{[s]_\rq!}{[\ell]_\rq![s-\ell]_\rq!}.
\label{cs}\ee
These are exactly the same as in \cite{Lusztig}. But from now on,
we will use the cyclic property as in \cite{DFM}.
We let $s=kN+p$ for $0\le k\le \lfloor m/N\rfloor$, and $0\le p\le p_k$, where for
$0\le k\le \lfloor m/N\rfloor-1$, $p_k=N-1$, but $p_{ \lfloor m/N\rfloor}=N\{ m/N\}$,
with $\{ x\}$ the fractional part of $x$. Using (3.55) of \cite{DFM}, namely
\be 
\sfactor s\ell_\rq=\sfactor{kN+p}\ell_\rq=\rq^{kN\ell}\sfactor p\ell_\rq\label{kNp},
\ee
we rewrite $c_s$ in (\ref{cs}) as
\be c_{kN+p}=(-1)^{m-kN-p}\rq^{(m-kN-p)(2n-m+1)}
\sum_{\ell=0}^{m-2n-1}(-1)^{\ell} \rq^{\ell(1-p)}\sfactor p{\ell}_\rq.
\label{cjNp} \ee
For $p\le {m-2n-1}$, we have
\ba
\sum_{\ell=0}^{m-2n-1}(-1)^{\ell} \rq^{\ell(1-p)}
 \sfactor p{\ell}_\rq=\sum_{\ell=0}^{p}(-1)^{\ell} \rq^{\ell(1-p)}
 \sfactor p{\ell}_\rq=\delta_{p,0},\ea
where 1.3.4 of \cite{Lusztig}, or (3.58) of \cite{DFM}, is used. 
Consequently, (\ref{gh}) can be rewritten as
 \ba\fl
 g=\sum_{k=0}^{\lfloor m/N\rfloor}\sum_{p=0}^{p_k} c_{kN+p}\,
 \theta_i^{(m-kN-p)}\theta_j^{(n)}\theta_i^{(kN+p)}\nonumber\\
 \fl=\sum_{k=0}^{\lfloor m/N\rfloor} c_{kN}\,
 \theta_i^{(m-kN)}\theta_j^{(n)}\theta_i^{(kN)}+
 \sum_{k=0}^{\lfloor m/N\rfloor}\sum_{p=m-2n}^{p_k} c_{kN+p}\,
 \theta_i^{(m-kN-p)}\theta_j^{(n)}\theta_i^{(kN+p)}.
\label{id1} \ea
Since
\be
\theta_i^{(kN+p)}\theta_i^{(N-m+2n)}=
\sfactor{kN+N+p-m+2n}{N-m+2n}_\rq\theta_i^{(kN+N+p-m+2n)},\ee
and, for $m-2n\le p\le N-1$,
\be
\sfactor{kN+N+p-m+2n}{N-m+2n}_\rq=
\rq^{(N-m+2n)N(k+1)}\sfactor{p-m+2n}{N-m+2n}_\rq=0,
\ee
we find
\be
\sum_{p=m-2n}^{p_k} c_{kN+p}\,
 \theta_i^{(m-kN-p)}\theta_j^{(n)}\theta_i^{(kN+p)}\theta_i^{(N-m+2n)}=0.
 \ee
Thus by multiplying $\theta_i^{(N-m+2n)}$ to $g$, we can get rid of
the second term in (\ref{id1}), or
\ba\fl
0=g\theta_i^{(N-m+2n)}=\sum_{k=0}^{\lfloor m/N\rfloor} c_{kN} \theta_i^{(m-kN)}\theta_j^{(n)}
\theta_i^{(kN)}\theta_i^{(N-m+2n)}\nonumber\\
=\sum_{k=0}^{\lfloor m/N\rfloor} c_{kN} \theta_i^{(m-kN)}\theta_j^{(n)}
\theta_i^{(kN+N-m+2n)}\sfactor{kN+N-m+2n}{N-m+2n}_\rq\nonumber\\
=(-1)^{m}\rq^{m(2n-m+1)} \Bigg\{\sum_{k=0}^{\lfloor m/N\rfloor} (-1)^{k}\theta_i^{(m-kN)}\theta_j^{(n)}
\theta_i^{(kN+N-m+2n)}\Bigg\}.
\label{id2}\ea
If we let $n=Q$ and $m=N+Q$, then (\ref{id2}) becomes
\ba
\theta_i^{(N+Q)}\theta_j^{(Q)}\theta_i^{(Q)}=
 \theta_i^{(Q)}\theta_j^{(Q)}\theta_i^{(N+Q)}.
\label{thetaij} \ea
Now letting $n=N+Q$ and $m=3N+Q$ in (\ref{id2}), we find
\ba\fl
\theta_i^{(3N+Q)}\theta_j^{(N+Q)}\theta_i^{(Q)}&-
\theta_i^{(2N+Q)}\theta_j^{(N+Q)}\theta_i^{(N+Q)}
\nonumber\\
&+\theta_i^{(N+Q)} \theta_j^{(N+Q)}\theta_i^{(2N+Q)}-
 \theta_i^{(Q)}\theta_j^{(N+Q)}\theta_i^{(Q+3N)}=0.
\label{thetaiji} \ea

In order to show that (\ref{CBC1}) holds, we put $\theta_i={\bf E}_0$
and $\theta_j={\bf E}_1$ in (\ref{thetaij}), and then use (\ref{FE01})
and (\ref{BEn}). Similarly, from (\ref{thetaiji}) we obtain
\ba
\fl
 {\bf C}_0^{(3N+Q)}{\bf B}_1^{(N+Q)}{\bf C}_0^{(Q)}&-
{\bf C}_0^{(2N+Q)}{\bf B}_1^{(N+Q)}{\bf C}_0^{(N+Q)}
\nonumber\\
&+{\bf C}_0^{(N+Q)} {\bf B}_1^{(N+Q)}{\bf C}_0^{(2N+Q)}-
{\bf C}_0^{(Q)}{\bf B}_1^{(N+Q)}{\bf C}_0^{(Q+3N)}=0.
 \label{C0B1CB} \ea
Next we shall prove by induction the identity
\be
 {\bf C}_0^{(jN+Q)}{\bf B}_1^{(Q)}{\bf C}_0^{(Q)}=
 {\bf C}_0^{(Q)}{\bf B}_1^{(Q)}{\bf C}_0^{(jN+Q)}.
 \label{CBCj}\ee
For $j=1$, it is identical to (\ref{CBC1}). Assuming it holds for $j$,
we shall prove it for $j+1$. It is easy to verify that
 \be\fl
{\bf C}_0^{( jN+Q)}{\bf C}_0^{( kN)}=\sfactor{kN+jN+Q}{jN+Q}{\bf C}_0^{( jN+kN+Q)}
=\bino{k+j}{j}{\bf C}_0^{( jN+kN+Q)},
\label{mulo}\ee
so that for $k=1$, we have
\ba \fl
(1+j){\bf C}_0^{( jN+N+Q)}{\bf B}_1^{(Q)}{\bf C}_0^{(Q)}
={\bf C}_0^{(N)}{\bf C}_0^{( jN+Q)}{\bf B}_1^{(Q)}{\bf C}_0^{(Q)}
\nonumber\\
={\bf C}_0^{(N)}{\bf C}_0^{(Q)}{\bf B}_1^{(Q)}{\bf C}_0^{( jN+Q)}
={\bf C}_0^{(N+Q)}{\bf B}_1^{(Q)}{\bf C}_0^{(Q)}{\bf C}_0^{( jN)}
\nonumber\\
={\bf C}_0^{(Q)}{\bf B}_1^{(Q)}{\bf C}_0^{(N+Q)}{\bf C}_0^{( jN)}
=(1+j){\bf C}_0^{(Q)}{\bf B}_1^{(Q)}{\bf C}_0^{( jN+N+Q)}.
\ea
Thus we have proven (\ref{CBCj}) hold for any $j$. Furthermore, we can also show
\ba\fl
 {\bf C}_0^{(jN+Q)}{\bf B}_1^{(Q)}{\bf C}_0^{(kN+Q)}=
 {\bf C}_0^{(jN+Q)}{\bf B}_1^{(Q)}{\bf C}_0^{(Q)}{\bf C}_0^{(kN)}
 ={\bf C}_0^{(Q)}{\bf B}_1^{(Q)}{\bf C}_0^{(jN+Q)}{\bf C}_0^{(kN)}
 \nonumber\\
 \fl\qquad
 =\bino{k+j}{j}{\bf C}_0^{(Q)}{\bf B}_1^{(Q)}{\bf C}_0^{( jN+kN+Q)}
 =\bino{k+j}{j}{\bf C}_0^{( jN+kN+Q)}{\bf B}_1^{(Q)}{\bf C}_0^{(Q)}.
 \label{CBCjk}\ea
 We then can use these formulae to move things around,  for example,
\ba
{\bf x}^{-}_{1,Q}({\bf x}^{+}_{0,Q})^3&={\bf C}_0^{(Q)}{\bf B}_1^{(N+Q)}{\bf C}_0^{(N+Q)}
{\bf B}_1^{(Q)}
{\bf C}_0^{(N+Q)}{\bf B}_1^{(Q)}{\bf C}_0^{(N+Q)}{\bf B}_1^{(Q)}\nonumber\\
&=6{{\bf C}_0^{(Q)}{\bf B}_1^{(N+Q)}{\bf C}_0^{(3N+Q)}}{\bf B}_1^{(Q)}
{\bf C}_0^{(Q)}{\bf B}_1^{(Q)}{\bf C}_0^{(Q)}{\bf B}_1^{(Q)}.
\ea
Similarly, we can show
\ba
\fl
{\bf x}^{+}_{0,Q}{\bf x}^{-}_{1,Q}({\bf x}^{+}_{0,Q})^2
=2{{\bf C}_0^{(N+Q)}{\bf B}_1^{(N+Q)}{\bf C}_0^{(2N+Q)}}
{\bf B}_1^{(Q)}{\bf C}_0^{(Q)}{\bf B}_1^{(Q)}{\bf C}_0^{(Q)}{\bf B}_1^{(Q)},
\nonumber\\
\fl
({\bf x}^{+}_{0,Q})^2{\bf x}^{-}_{1,Q}{\bf x}^{+}_{0,Q}
=2{{\bf C}_0^{(2N+Q)}{\bf B}_1^{(N+Q)}{\bf C}_0^{(N+Q)}}
{\bf B}_1^{(Q)}
{\bf C}_0^{(Q)}{\bf B}_1^{(Q)}{\bf C}_0^{(Q)}{\bf B}_1^{(Q)},
\nonumber\\
\fl
({\bf x}^{+}_{0,Q})^3{\bf x}^{-}_{1,Q}
=6{{\bf C}_0^{(3N+Q)}{\bf B}_1^{(N+Q)}{\bf C}_0^{(Q)}}
{\bf B}_1^{(Q)}{\bf C}_0^{(Q)}{\bf B}_1^{(Q)}{\bf C}_0^{(Q)}{\bf B}_1^{(Q)},
\ea
so that
\ba
\fl
[[[{\bf x}^{-}_{1,Q},{\bf x}^{+}_{0,Q}],{\bf x}^{+}_{0,Q}],{\bf x}^{+}_{0,Q}]\nonumber\\
\fl\quad
={\bf x}^{-}_{1,Q}({\bf x}^{+}_{0,Q})^3
-3({\bf x}^{+}_{0,Q}){\bf x}^{-}_{1,Q}({\bf x}^{+}_{0,Q})^2
+3({\bf x}^{+}_{0,Q})^2({\bf x}^{-}_{1,Q}{\bf x}^{+}_{0,Q})
-({\bf x}^{+}_{0,Q})^3{\bf x}^{-}_{1,Q}\nonumber\\
\fl\quad
=6\bigg[{\bf C}_0^{(Q)}{\bf B}_1^{(N+Q)}{\bf C}_0^{(3N+Q)}
-{\bf C}_0^{(N+Q)}{\bf B}_1^{(N+Q)}{\bf C}_0^{(2N+Q)}
+{\bf C}_0^{(2N+Q)}{\bf B}_1^{(N+Q)}{\bf C}_0^{(N+Q)}\nonumber\\
\fl\hspace{0.87in}
-{\bf C}_0^{(3N+Q)}{\bf B}_1^{(N+Q)}{\bf C}_0^{(Q)}\bigg]
{\bf B}_1^{(Q)}
{\bf C}_0^{(Q)}{\bf B}_1^{(Q)}
{\bf C}_0^{(Q)}{\bf B}_1^{(Q)}=0.
\ea 
Here (\ref{C0B1CB}) has been used. Likewise, by different choices of
the $\theta_i$, we can prove
\ba
&[[[{\bf x}^{+}_{0,Q},{\bf x}^{-}_{1,Q}],{\bf x}^{-}_{1,Q}],{\bf x}^{-}_{1,Q}]=0,\quad
[[[{\bf\bar x}^{-}_{0,Q},{\bf \bar x}^{+}_{-1,Q}],
  {\bf\bar x}^{+}_{-1,Q}],{\bf \bar x}^{+}_{-1,Q}]=0,\nonumber\\
&
[[[{\bf \bar x}^{+}_{-1,Q},{\bf\bar x}^{-}_{0,Q}],
  {\bf \bar x}^{-}_{0,Q}],{\bf\bar x}^{-}_{0,Q}]=0.
\ea
Thus we prove the Serre relations for generators of the loop algebra.
\subsection{Summary}
The weights of the CSOS models given by (\ref{csos2j}) for $\ell=2$ satisfy
the Yang--Baxter equations (\ref{ybeUU}). As a consequence, we were able to
show that the eigenvalues of the corresponding transfer matrix $\btau_{2,j}$ are
given by (\ref{tau2}), in which $F(t)$ is a polynomial of degree $R$ given
in (\ref{defF}), with roots $x_i$ for $i=1,\cdots,R$ satisfying the
Bethe Ansatz equations (\ref{bethe}). We then used the functional
relations (\ref{funljp}) to show that the eigenvalues of the transfer
matrices $\btau_{\ell,j}$ of the CSOS model are given by (\ref{tauljt}).

Substituting this result in the functional relations (\ref{fun2}) for the
product of two transfer matrix eigenvalues, we found that these
reduce to (\ref{fun3}) with the same polynomial ${\cal P}$ independent of
$\ell$. We then examined the various properties of these eigenvalues,
enabling us to show that they are given by (\ref{calTxy}) and (\ref{hatTyx2}).
The polynomial ${\cal P}(t_q^N/t_p^N)$ in (\ref{fun3}) and (\ref{Bamp}) is
a polynomial in $t_q^N$ of degree $m_E$, and for each root of ${\cal P}$,
there are two choices of $\lambda_q$, as can be seen from (\ref{xyt}) and
(\ref{PGG}). This shows that there are $2^{m_E}$ possible eigenvalues
of the transfer matrix associated with the polynomial $F(t)$.

Since the transfer matrix ${\cal T}(x_q,y_q)$ and $\btau_{\ell,j}$ commute
with $\btau_{2,j}$, they have the same eigenvectors. To each $F(t)$,
corresponding to one eigenvalue (\ref{tau2}) of $\btau_{2,j}$, there
are $2^{m_E}$ different eigenvalues of ${\cal T}(x_q,y_q)$.
This means that the eigenspace associated with this eigenvalue of $\btau_{2,j}$
has a $2^{m_E}$-fold degeneracy. This clearly points to the existence
of the quantum loop algebra in the CSOS model derived from $\btau_{2,j}$.
The transfer matrices $\btau_{\ell,j}$ of CSOS models with weights
(\ref{squaresos}) were shown to have eigenvalues given by (\ref{tauljt}).

From (\ref{Bamp}), we can see that $m_E=\lfloor \big(jL-L-2R-P_a-P_b\big)/N\rfloor$.
The $2^{m_E}$-fold degeneracy in the $\btau_{2,j}$ model was verified by finite-size
calculations, with a few exceptions. As an example, we have found an eigenvalue
of $\btau_{2,j}$ associated with a polynomial $F(t)$ of degree $R=3$ for the case
of $N=3$, $L=6$, $j=2$ and $Q=1$, for which $m_E$ is negative. To understand
this anomaly, we have calculated the eigenvalue of ${\cal T}(x_q,y_q)$ and
${\hat{\cal T}}(y_q,x_q)$ and found ${\hat{\cal T}}(y_q,x_q)=0$ for that case,
so that (\ref{fun3}) still holds.

In Section 4, we first showed that the leading coefficients of the monodromy
matrix of the CSOS models, ${\bf C}_{L-1}$, ${\bf B}_{L}$, ${\bf C}_{0}$
and ${\bf B}_{1}$ are related to a $j^L$-dimensional representation of the
affine quantum group $U_{\rm q}(\widehat{\mathfrak{sl}}_2)$. We then
showed that the resulting generators of the loop algebra given by
(\ref{generators}) indeed satisfy the Serre relations.

\section*{Acknowledgment}
The authors thank Professors Rodney Baxter, Vladimir Bazhanov, Murray Batchelor,
and Vladimir Mangazeev for much encouragement and hospitality and the Australian National University, the Australian Research Council and other Australian sources
for financial support during the several years that the research reported here was done.
They also thank Professor Xi-Wen Guan, the Wuhan Institute of Physics and Mathematics
and the Chinese Academy of Science for financial support and hospitality during a
three month visit in Wuhan during the summer of 2014, and another visit to Beijing in 2015.
They thank Dr. Andreas Kl\"umper for showing the simple derivation of the
functional relations of the $T$-system and for providing relevant references.
Finally, early work on the Serre relations in the superintegrable case has been
supported in part by the National Science Foundation under grant No. PHY-07-58139.

\appendix
\section{Decomposition of a square}
\subsection{The square weight $U(a,b,c,d)$}
Consider the square resulting from the star-weight (\ref{square}) and let 
\be
(x_{q'},y_{q'},\mu_{q'})=(y_q,\omega^\ell x_q,\mu_q^{-1}).
\label{qq'}\ee
Then the four Boltzmann weights in this square are given by (\ref{weights})
and are re-expressed in terms of the $\omega$-Pochhammer symbol
(sometimes called the $\omega$-shifted factorial)
\be
(x;\omega)_n=\prod_{m=0}^{n-1}(1-x\omega^m),\quad
 (x;\omega)_{-n}=\frac 1{(\omega^{-n}x;\omega)_n},
\label{Poch}\ee
as
\ba
W_{pq}(a-e)=\Bigg[{\mu_p y_q\over\mu_q y_p}\Bigg]^{a-e}
\frac{(\omega x_p/y_q;\omega)_{a-e}}{(\omega x_q/y_p;\omega)_{a-e}},
\nonumber\\
 \overline W_{p'q}(b-e)=
 \Bigg[\frac{\omega\mu_{p'} x_{p'}\mu_q}  {y_q}\Bigg]^{b-e}
 \frac{(x_q /x_{p'};\omega)_{b-e}}{(\omega y_{p'}/ y_q;\omega)_{b-e}},
 \nonumber\\
   \overline W_{pq'}(e-d)=\Bigg[{\mu_p y_q\over\mu_q y_p}\Bigg]^{e-d}
\frac{(\omega^{1-e+d} x_p/y_q;\omega)_{e-d}}
{(\omega^{\ell-e+d} x_q/y_p;\omega)_{e-d}},
\nonumber\\
W_{p'q'}(e-c)=\Bigg[\frac{\mu_{p'} x_{p'}\mu_q}  {y_q}\Bigg]^{e-c}
\frac{(\omega^{\ell-e+c}x_q/ x_{p'};\omega)_{e-c}}
{(\omega^{-e+c}y_{p'}/ y_q;\omega)_{e-c}}. 
 \label{weightsqq'}\ea
Using the relation
\be ( x;\omega)_a(\omega^a x;\omega)_{b}=( x;\omega)_{a+b},
\label{poch2}\ee
we may combine these products as
\ba
\fl
(\omega^{1-e+d} x_p/y_q;\omega)_{e-d}(\omega x_p/y_q;\omega)_{a-e}
=(\omega^{1-e+d} x_p/y_q;\omega)_{a-d},
\nonumber\\
\fl
(\omega^{-e+c}y_{p'}/ y_q;\omega)_{e-c}(\omega y_{p'}/ y_q;\omega)_{b-e}
=(\omega^{-e+c}y_{p'}/ y_q;\omega)_{b-c+1}/(1-y_{p'}/ y_q),
\nonumber\\
\fl
{(\omega x_q/y_{p};\omega)_{\ell-1}}\Big/{(\omega x_q/y_p;\omega)_{a-e}}
={(\omega^{1-e+a} x_q/y_{p};\omega)_{\ell-1-a+e}}
\nonumber\\
=(\omega^{1-e+a}x_q/y_{p};\omega)_{\ell-1-a+d}
{(\omega^{\ell-e+d} x_q/y_p;\omega)_{e-d}},
\nonumber\\
\fl
(\omega^{\ell-e+c}x_q/ x_{p'};\omega)_{e-c}
=(\omega^{\ell-e+c}x_q/ x_{p'};\omega)_{e-c-\ell}
(x_q /x_{p'};\omega)_{\ell},
\nonumber\\
\fl
(\omega^{\ell-e+c}x_q/ x_{p'};\omega)_{e-c-\ell}(x_q /x_{p'};\omega)_{b-e}
=(\omega^{\ell-e+c}x_q/ x_{p'};\omega)_{b-c-\ell},
\ea
so that the star-weight (\ref{square}) can be rewritten as
\ba
\fl
U_{pp'qq'}(a,b,c,d)=\Bigg[{\mu_p y_q\over\mu_q y_p}\Bigg]^{\alpha}
\Bigg[\frac{\mu_{p'} x_{p'}\mu_q}  {y_q}\Bigg]^{\beta} 
\frac{(1-y_{p'}/ y_q)(x_q /x_{p'};\omega)_{\ell}}{(\omega x_q/y_{p};\omega)_{\ell-1}}
\sum_{e=1}^N\omega^{b-e}J,
\label{square1}\ea
where $\alpha=a-d$, $\beta=b-c$, 
and
\be
\fl
J=\frac{
(\omega^{1-e+d}x_p/y_{q};\omega)_{\alpha}
(\omega^{1-e+a}x_q/y_{p};\omega)_{\ell-1-\alpha}
(\omega^{\ell-e+c}x_q/x_{p'};\omega)_{\beta-\ell}}
{(\omega^{c-e}y_{p'}/y_{q};\omega)_{1+\beta}}.
\label{defJ}\ee
\subsection{$U_{pp'qq'}(a,b,c,d)$ for $0\le\alpha\le\ell-1$ and $\ell\le\beta\le N-1$}
Now consider the case $0\le\alpha\le\ell-1$ and $\ell\le\beta\le N-1$. We use
\be
(\omega^a x;\omega)_N=(1-x^N)=(\omega^a x;\omega)_b(\omega^{a+b} x;\omega)_{N-b},
\label{flip}\ee 
to flip the denominator in $J$ so that 
\ba\fl
J=
(\omega^{1-e+d}x_p/y_{q};\omega)_{\alpha}
(\omega^{1-e+a}x_q/y_{p};\omega)_{\ell-1-\alpha}
\nonumber\\
\times\,(\omega^{\ell-e+c}x_q/x_{p'};\omega)_{\beta-\ell}
(\omega^{1-e+b}y_{p'}/y_{q};\omega)_{N-1-\beta}\Big/(1-y^N_{p'}/y^N_{q}).
\label{J1}\ea 
Using the function $\Phi$ introduced in (BBP:3.26)
\ba
(\omega ux;\omega)_{\alpha}(\omega uy;\omega)_{\beta}
=\sum_{n=0}^{\alpha+\beta}u^n\Phi(x,y)_n^{\alpha,\beta},
\label{Phi}\ea
we find
\ba\fl
J=\frac1{1-y^N_{p'}/y^N_{q}}
\sum_{m=0}^{\ell-1}\sum_{n=0}^{N-\ell-1}
\Bigg[\frac{\omega^{d-e}x_p}{y_{q}}\Bigg]^{m}
\Bigg[\frac{\omega^{b-e}y_{p'}}{y_{q}}\Bigg]^{n}
\nonumber\\
\times\;\Phi\Bigg(1,\frac{\omega^\alpha t_q}{t_{p}}\Bigg)_{m}^{\alpha,\ell-\alpha-1}
\Phi\Bigg(1,\frac{\omega^{\ell-1-\beta} t_q}{t_{p'}}\Bigg)_{n}^{N-1-\beta,\beta-\ell}.
\label{JJ1}\ea 
It is easily seen that $0\le n+m\le N-2$. Substituting (\ref{JJ1}) into (\ref{square1}),
we find that the summation over $e$ yields 
\be
U_{pp'qq'}(a,b,c,d)=0,  \quad \hbox {for $0\le\alpha\le \ell-1$ and $\ell\le\beta\le N-1$}.
\label{square2}\ee
This is the result (BBP:3.17) in \cite{BBP}. 
\subsection{Properties of $\Phi\big(x, y\big)_n^{\alpha,\beta}$ }
We shall now express the function $\Phi$ defined in (\ref{Phi}) in terms
of basic hypergeometric series and explore some of its properties.
From corollary 10.2.2(c) in \cite{AAR}, we have
\be\fl
\sum_{k=0}^n\left[n\atop k\right](-1)^k\omega^{\halfs k(k-1)} x^k
=(x;\omega)_n,\quad
\left[n\atop k\right]=\frac{[n]!}{[n-k]![k]!}
=\frac{(\omega^{1+n-k};\omega)_k}{(\omega;\omega)_k}.
\label{coro10c}\ee
Consequently, (\ref{Phi}) becomes
\be\fl
(\omega ux;\omega)_{\alpha}(\omega uy;\omega)_{\beta}
=\sum_{k=0}^{\alpha}\sum_{k'=0}^{\beta}\Bigg[{\alpha\atop k}\Bigg]
\Bigg[{\beta\atop k'}\Bigg](-1)^{k+k'}u^{k+k'}
\omega^{\halfs k(k+1)+\halfs k'(k'+1)}x^k y^{k'}.
\label{Phi2}\ee
Letting $n=k+k'$, we find 
\ba
\Phi(x,y)_n^{\alpha,\beta}=(-1)^n\omega^{\halfs n(n+1)}
\sum_{k=0}^n
\Bigg[{\alpha\atop k}\Bigg]\Bigg[{\beta\atop n-k}\Bigg]
\omega^{k(k-n)}y^n(x/y)^k
\nonumber\\
\hspace{50pt}
=(-1)^n\omega^{\halfs n(n+1)}\sum_{k=0}^n
\Bigg[{\alpha\atop n-k}\Bigg]\Bigg[{\beta\atop k}
\Bigg]\omega^{k(k-n)}x^n(y/x)^k.
\label{Phi3}\ea
We use (\ref{coro10c}), (\ref{poch2}) and
\be
(x;\omega)_n=(-1)^n\omega^{\halfs n(n-1)}x^n(\omega^{1-n}/x,\omega)_n,
\label{poch3}\ee
to write
\ba
\Bigg[{\alpha\atop k}\Bigg]&
=\frac{(\omega^{1+\alpha-k};\omega)_k}{(\omega;\omega)_k}
=(-1)^k\omega^{\alpha k-\halfs k(k-1)}
\frac{(\omega^{-\alpha};\omega)_k}{(\omega;\omega)_k},
\nonumber\\
\Bigg[{\alpha\atop n-k}\Bigg]&=(-1)^k\omega^{n k-\halfs k(k-1)}
\frac{(\omega^{1+\alpha-n};\omega)_n(\omega^{-n},\omega)_k}
{(\omega;\omega)_n(\omega^{1+\alpha-n};\omega)_k},
\label{qbino}\ea
so that
\ba\fl
(-1)^n\omega^{-\halfs n(n+1)}\Phi(x,y)_n^{\alpha,\beta}&=
\frac{(\omega^{1+\beta-n};\omega)_n y^n}{(\omega;\omega)_n}
\hypp{1}{\omega^{-n}}{\omega^{-\alpha}}{\omega^{1+\beta-n}}
{\frac{x\omega^{\alpha+1}}y}
\label{phitohy.a}\\
\fl &=\frac{(\omega^{1+\alpha-n};\omega)_n x^n}{(\omega;\omega)_n}
\hypp{1}{\omega^{-n}}{\omega^{-\beta}}{\omega^{1+\alpha-n}}
{\frac{y\omega^{\beta+1}}x},
\label{phitohy.b}\ea
with basic hypergeometric function ${}_2\Phi_1$.
Particularly for $\beta=\ell-\alpha-1$, we have
\ba\fl
\Phi(1,\omega^{\alpha} y)_n^{\alpha,\ell-\alpha-1}&=(-1)^n\omega^{\halfs n(n+1)}
\frac{(\omega^{1+\alpha-n};\omega)_n}{(\omega;\omega)_n}
\hypp{1}{\omega^{-n}}{\omega^{1+\alpha-\ell}}{\omega^{1+\alpha-n}}{{\omega^\ell}y}\label{Phi5a}\\
\fl &=(-1)^n\omega^{\halfs n(n+1)+\alpha n}
\frac{(\omega^{\ell-\alpha-n};\omega)_n y^n}{(\omega;\omega)_n}
\hypp{1}{\omega^{-n}}{\omega^{-\alpha}}{\omega^{\ell-\alpha-n}}{\frac{\omega}y}.
\label{Phi5}\ea
From (\ref{Phi}) or (\ref{Phi2}), we see that, when $0\le\alpha\le\ell-1$,
$\Phi(1,\omega^{\alpha} y)_n^{\alpha,\ell-\alpha-1}=0$
for $n\ge\ell$. Since the basic hypergeometric function in (\ref{Phi5})
is symmetric in $n$ and $\alpha$, we find
\ba\fl
\Phi(1,\omega^{n} y)_\alpha^{n,\ell-n-1}
=&(-y)^{\alpha-n}\omega^{\halfs \alpha(\alpha+1)-\halfs n(n+1)}
\nonumber\\
\fl&\times\,\frac{(\omega^{\ell-\alpha-n};\omega)_\alpha (\omega;\omega)_n}
{(\omega^{\ell-\alpha-n};\omega)_n(\omega;\omega)_\alpha}
\Phi(1,\omega^{\alpha} y)_n^{\alpha,\ell-\alpha-1}.
\label{Phi6.a}\ea
Now we use (\ref{poch2}) and then (\ref{poch3}) to write
\ba\fl
\frac{(\omega^{\ell-\alpha-n};\omega)_\alpha}
{(\omega^{\ell-\alpha-n};\omega)_n}
=(\omega^{\ell-\alpha};\omega)_{\alpha-n}=
(-1)^{\alpha-n}\omega^{\ell(\alpha-n)-\halfs (\alpha-n)(\alpha+n+1)}
(\omega^{1+n-\ell};\omega)_{\alpha-n},
\nonumber\\
\fl
(\omega^{1-\ell+n};\omega)_{\alpha-n}=\frac{(\omega^{1-\ell};\omega)_\alpha}
{(\omega^{1-\ell};\omega)_n}=\frac{(\omega;\omega)_{N-\ell+\alpha}}
{(\omega;\omega)_{N-\ell+n}}
=\frac{(\omega;\omega)_{\alpha}
(\omega^{1+\alpha};\omega)_{N-\ell}}
{(\omega;\omega)_{n}(\omega^{1+n};\omega)_{N-\ell}},
\ea
so that (\ref{Phi6.a}) can be further simplified to
\be
\Phi(1,\omega^{n} y)_\alpha^{n,\ell-n-1}
=\big(\omega^\ell y\big)^{\alpha-n}\frac{(\omega^{1+\alpha};\omega)_{N-\ell}}
{(\omega^{1+n};\omega)_{N-\ell}}\,
\Phi(1,\omega^{\alpha} y)_n^{\alpha,\ell-\alpha-1}.
\label{Phi6}\ee
\subsection{Identity (3.33) in \cite{BBP}}
The identity (BBP:3.33), which was proven using the star-triangle
equations of the chiral Potts weights, can be rewritten as
\ba\fl
\Phi(1,\omega^{\ell-\beta-1}t)_{N-m-1}^{N-\beta-1,N+\beta-\ell}
=(\omega^{\ell }t)^{\beta-m}
(t\omega^{\ell};\omega)_{N-\ell}
\frac{(\omega^{1+\beta};\omega)_{N-\ell}}
{(\omega^{1+m};\omega)_{N-\ell}}
\Phi(1,\omega^{\beta} t)_m^{\beta,\ell-\beta-1}.
\label{BBP333}\ea
Here we shall present a different proof.

For the root-of-unity case, Theorem 10.2.1 in \cite{AAR} does not hold.
In fact, following their method, we find instead
\be
\sum_{n=0}^{N-1}\frac{(a;\omega)_n} {(\omega;\omega)_n} x^n
=N^{-1}\sum_{n=0}^{N-1}{(\omega^{-n}x;\omega)_n} {(ax;\omega)_{N-n-1}}.
\ee
Since the proof of Theorem 10.10.1 in \cite{AAR} is based on
Theorem 10.2.1, it is not valid for $q^N=1$. It needs to be modified to\\
\textbf{Theorem 10.10.1 for \boldmath{$\omega^N=1$}}
\be\fl
\hypp{1}{\omega^{\alpha}}{\omega^{\beta}}{\omega^{\gamma}}{t}=
\big(\omega^{\alpha+\beta-\gamma}t;\omega\big)^{\vp}_{N-\alpha-\beta+\gamma}\,\,
\hypp{1}{\omega^{\gamma-\alpha}}{\omega^{\gamma-\beta}}{\omega^{\gamma}}{\omega^{\alpha+\beta-\gamma}t}.
\label{coro10}\ee
Before applying this, we first use (\ref{phitohy.b}) to derive
\ba\fl
\Phi(1,\omega^{\ell-\beta-1}t)_{N-m-1}^{N-\beta-1,N+\beta-\ell}
\nonumber\\
=(-1)^m\omega^{\halfs m(m+1)}
\frac{(\omega^{1+m-\beta};\omega)_{N-m-1} }{(\omega;\omega)_{N-m-1} }\,\,
\hypp{1}{\omega^{m+1}}{\omega^{\ell-\beta}}{\omega^{1+m-\beta}}{t}.
\label{proof1a}\ea
Next we use (\ref{coro10}) and then (\ref{Phi5a}) to find
\ba\fl
\hypp{1}{\omega^{m+1}}{\omega^{\ell-\beta}}{\omega^{1+m-\beta}}{t}=
\big(\omega^{\ell}t;\omega\big)^{\vp}_{N-\ell}\,\,
\hypp{1}{\omega^{-\beta}}{\omega^{1+m-\ell}}{\omega^{1+m-\beta}}{\omega^{\ell}t}
\nonumber\\
=\big(\omega^{\ell}t;\omega\big)^{\vp}_{N-\ell}\,\,\frac{(\omega;\omega)_{\beta}}{(\omega^{1+m-\beta};\omega)_{\beta} }
(-1)^\beta\omega^{-\halfs\beta(\beta+1)}\Phi(1,\omega^{m}t)_{\beta}^{m,\ell-m-1}.
\label{proof1}\ea
Using (\ref{poch2}), followed by (\ref{poch3}) for the numerator, we can write
\ba\fl
\frac{(\omega^{1+m-\beta};\omega)_{N-m-1} (\omega;\omega)_{\beta}}
{(\omega^{1+m-\beta};\omega)_{\beta} (\omega;\omega)_{N-m-1} }=
\frac{(\omega^{1+m};\omega)_{N-m-1-\beta} }{(\omega^{1+\beta};\omega)_{N-m-1-\beta}}\nonumber\\
\fl
=(-1)^{N-m-\beta-1}\omega^{(N-m-\beta-1)[\halfs(N-m-\beta-2)+m+1]}
=(-1)^{m-\beta}\omega^{-\halfs(m+\beta+1)(m-\beta)}.
\label{ratio}\ea
Substituting (\ref{proof1}) into (\ref{proof1a}) and using (\ref{ratio}),
we simplify (\ref{proof1a}) to
\be
\Phi(1,\omega^{\ell-\beta-1}t)_{N-m-1}^{N-\beta-1,N+\beta-\ell}=
\big(\omega^{\ell}t;\omega\big)^{\vp}_{N-\ell}\,
\Phi(1,\omega^{m}t)_{\beta}^{m,\ell-m-1}.
\label{BBP333b}\ee
Finally we use (\ref{Phi6}) with $n=m$ and $\alpha=\beta$
to find that (\ref{BBP333b}) becomes (\ref{BBP333}).
\subsection{$U_{pp'qq'}(a,b,c,d)$ for $0\le\alpha,\beta\le\ell-1$.}
We see from (\ref{square2}), that the square becomes block triangular.
We shall first calculate the upper diagonal block. For $\beta-\ell\le 0$
and (\ref{Poch}) we find
\ba\fl
(\omega^{\ell-e+c}x_q/x_{p'};\omega)_{\beta-\ell}
&=1/(\omega^{b-e}x_q/x_{p'};\omega)_{\ell-\beta}
\nonumber\\
&=(\omega^{{\ell-e+c}}x_q/x_{p'};\omega)_{N-\ell+\beta}\Big/(1-x^N_q/x_{p'}^N).
\ea
Consequently, (\ref{J1}) becomes
\ba\fl
J=(1-y^N_{p'}/y^N_{q})^{-1}(1-x^N_q/x^N_{p'})^{-1}
(\omega^{1-e+d}x_p/y_{q};\omega)_{\alpha}
(\omega^{1-e+a}x_q/y_{p};\omega)_{\ell-1-\alpha}
\nonumber\\
\times\,(\omega^{1-e+b}y_{p'}/y_{q};\omega)_{N-1-\beta}
(\omega^{\ell-e+c}x_q/x_{p'};\omega)_{N+\beta-\ell}
\nonumber\\
\fl\quad
=(1-y^N_{p'}/y^N_{q})^{-1}(1-x^N_q/x^N_{p'})^{-1}
\sum_{m=0}^{\ell-1}\sum_{n=0}^{2N-\ell-1}\omega^{-ne-me}
\bigg(\frac{\omega^dx_p}{y_{q}}\bigg)^{m}
\bigg(\frac{\omega^b y_{p'}}{y_{q}}\bigg)^{n}
\nonumber\\
\times\,\Phi\bigg(1,\frac{\omega^\alpha t_q}{t_{p}}\bigg)_m^{\alpha,\ell-\alpha-1}
\Phi\bigg(1,\omega^{\ell-1-\beta}\frac{ t_q}{t_{p'}}\bigg)_n
^{N-1-\beta,N+\beta-\ell}.
\label{J2}\ea 
Substituting it into (\ref{square1}) and using (\ref{omega}), we find,
for $0\le \alpha,\beta\le \ell-1$,
\ba\fl
U_{pp'qq'}(a,b,c,d)=\Big(\frac{\mu_py_q}{\mu_qy_p}\Big)^{\alpha}
\Big(\frac{\mu_q\mu_{p'}x_{p'}}{y_q}\Big)^{\beta}
\frac{N\Omega_{pp'q}}{(1-x^N_q/x^N_{p'})}
\nonumber\\
\fl\qquad\times
\sum_{m=0}^{\ell-1}\omega^{(d-b)m}
\bigg(\frac{x_p}{y_{p'}}\bigg)^{m}
\Phi\bigg(1,\frac{\omega^\alpha t_q}{t_{p}}\bigg)_m^{\alpha,\ell-\alpha-1}
\Phi\bigg(1,\omega^{\ell-1-\beta}\frac{ t_q}{t_{p'}}\bigg)_{N-1-m}
^{N-1-\beta,N+\beta-\ell},
\label{square3}\ea
where
\be
\Omega_{pp'q}=\frac{(1-y_q/y_{p'})(x_q /x_{p'};\omega)_{\ell}}
{(1-y^N_{q}/y^N_{p'})(\omega x_q/y_{p};\omega)_{\ell-1}}.
\label{omega}\ee
Defining the function $F_{pq}$ as in \cite{BBP},
\be
F_{pq}(\ell,\alpha,m)=\Big({\mu_p\over y_p}\Big)^{\alpha}{x_p}^{m}\,
\Phi\bigg(1,\frac{\omega^\alpha t_q}{t_{p}}\bigg)_m^{\alpha,\ell-\alpha-1},
\label{defFpq}\ee
we use (\ref{BBP333b}) to rewrite (\ref{square3}) as
\ba
U_{pp'qq'}(a,b,c,d)=A_{pp'q}\Big(\frac{y_q}{\mu_q}\Big)^{\alpha-\beta}
U^{(\ell)}_{pp'q}(a,b,c,d),
\label{square3b}\ea
where
\be\fl
A_{pp'q}=\frac{N\Omega_{pp'q}
\big(\omega^{\ell}t_q/t_{p'};\omega\big)^{\vp}_{N-\ell}}
{(1-x^N_q/x^N_{p'})}=
\frac{N(1-y_q/y_{p'})(x_q/x_{p'};\omega)_\ell
\big(\omega^{\ell}t_q/t_{p'};\omega\big)^{\vp}_{N-\ell}}
{(1-y_q^N/y_{p'}^N) (1-x^N_q/x^N_{p'})(\omega x_{q}/y_p;\omega)_{\ell-1}},
\label{App'q}\ee
and
\be
U^{(\ell)}_{pp'q}(a,b,c,d)=
\sum_{m=0}^{\ell-1}\omega^{(d-b)m}\mu_{p'}^{\beta-m}
F_{pq}(\ell,\alpha,m)F_{p'q}(\ell,m,\beta).
\label{square4a}\ee
It is more convenient to use (\ref{square4a}) to calculate the weights of the square.
However, for comparing the lower diagonal block with the upper one, we must use (\ref{BBP333}) in (\ref{square3}) as was done in \cite{BBP}, to find
\be
U^{(\ell)}_{pp'q}(a,b,c,d)=
\sum_{m=0}^{\ell-1}\omega^{(d-b)m}
F_{pq}(\ell,\alpha,m)F_{p'q}(\ell,\beta,m)\frac{\eta_{q,\ell,\beta}}{\eta_{q,\ell,m}},
\label{square4b}\ee
with
\be
\eta_{q,\ell,m}=(\omega^\ell t_q)^m(\omega^{1+m};\omega)_{N-\ell}.
\ee
\subsection{$U_{pp'qq'}(a,b,c,d)$ for $\ell\le\alpha\le N-1$ and $\ell\le\beta\le N-1$}
To calculate the lower diagonal block and to put it in the same form as the upper block,
it is necessary to use (\ref{poch3}) in (\ref{defJ}) and then use (\ref{flip}) to obtain
\ba\fl
J=\omega^{\ell(d-c)+2e-b-d}(t_p/t_q)^\alpha(t_q/t_{p'})^\beta(x_{p'}/x_q)^\ell
(x_q/y_p)^{\ell-1}(y_q/y_{p'})
\nonumber\\
\fl\qquad\times(1-y^N_{p}/x^N_q)^{-1}(1-y^N_{q}/y^N_{p'})^{-1}
(\omega^{e-a}y_{q}/x_p;\omega)_{\alpha}
(\omega^{1+e-d-\ell}y_{p}/x_q;\omega)_{N+\ell-1-\alpha}
\nonumber\\
\fl\qquad\times(\omega^{1+e-b}x_{p'}/x_q;\omega)_{\beta-\ell}
(\omega^{1+e-c}y_{q}/y_{p'};\omega)_{N-1-\beta}.
\label{J4}\ea
We again use (\ref{Phi}) and (\ref{omega}) to express (\ref{square1}) as
\begin{eqnarray}\fl
U_{pp'qq'}(a,b,c,d)=\bigg({\mu_px_p\over\mu_qx_q}\bigg)^{\alpha}
\bigg({\mu_{p'}x_q\mu_q\over y_{p'}}\bigg)^{\beta}
\frac{N\omega^{\ell(d-c+1)}(x_{p'}/y_p)^\ell\Omega_{pp'q}}{(1-x^N_q/y^N_{p})}
\nonumber\\
\fl
\qquad\times
\sum_{m=0}^{N-\ell-1}\omega^{(d-b+\ell)m}
\bigg(\frac{x_{p'}}{y_{p}}\bigg)^{m}
\Phi\bigg(1,\frac{\omega^\beta t_q}{t_{p'}}\bigg)_m^{\beta-\ell,N-\beta-1}
\Phi\bigg(1,\frac{ \omega^{\ell-1-\alpha}t_q}{t_{p}}\bigg)_{N-1-m}
^{N+\ell-1-\alpha,\alpha}.
\label{square5}\ea
Replacing $\ell\to N-\ell$, $\beta=\alpha-\ell$ and $t=\omega^\ell t_q/t_p$
in (\ref{BBP333}), we find
\ba\fl
\Phi\bigg(1,\frac{ \omega^{\ell-1-\alpha}t_q}{t_{p}}\bigg)_{N-1-m}
^{N+\ell-1-\alpha,\alpha}
=\big({t_q}/{t_{p}};\omega\big)^{\vp}_{\ell}
\Big(\frac{ t_q}{t_{p}}\Big)^{\alpha-\ell-m}
\nonumber\\
\times\,\frac{(\omega^{1+\alpha-\ell};\omega)_{\ell}}{(\omega^{1+m};\omega)_{\ell}}
\Phi\bigg(1,\frac{\omega^{\alpha} t_q}{t_{p}}\bigg)_m^{\alpha-\ell,N-\alpha-1}.
\ea
Using (\ref{defFpq}) and $t_{q'}=\omega^\ell t_q$, we may rewrite (\ref{square5}) as
\ba\fl
U_{pp'qq'}(a,b,c,d)={\hat A}_{pp'q}
(x_q\mu_q)^{\beta-\alpha}\omega^{\ell(d-c)}
{U}^{(N-\ell)}_{pp'q'}(a-\ell,b-\ell,c,d),
\label{square6a}\\
\fl
{U}^{(N-\ell)}_{pp'q'}(a-\ell,b-\ell,c,d)
\nonumber\\
\fl\qquad=\sum_{m=0}^{N-1-\ell}\omega^{(d-b+\ell)m}
F_{p'q'}(N-\ell,\beta-\ell,m)F_{pq'}(N-\ell,\alpha-\ell,m)
\frac{\eta_{q',N-\ell,\alpha-\ell}}{\eta_{q',N-\ell,m}},
\label{square6}\ea
where
\be
{\hat A}_{pp'q}=\bigg(\frac {\omega\mu_p\mu_{p'}x_px_{p'}}{y_py_{p'}}\bigg)^{\ell}
\frac{N\Omega_{pp'q}({t_q}/{t_{p}};\omega)^{\vp}_{\ell}}{(1-x^N_q/y^N_{p})}.
\label{hApp'q}\ee
\subsection{Functional relation}
For the case of cyclic boundary condition, $\sigma_{L+1}=\sigma_1$,
the product of two transfer matrices can be written as
\be
T_q{\hat T}_{q'}=
\prod_{i=1}^L U_{pp'qq'}(\sigma\vp_i,\sigma\vp_{i+1},\sigma'_{i+1},\sigma'_{i}).
\label{tqtq'}\ee
When the two rapidities $q$ and $q'$ are related by (\ref{qq'}), we find
from (\ref{square2}) that the squares $U_{pp'qq'}$ are block diagonal. Then
some of the factors in (\ref{square3b}) and (\ref{square6a}) cancel out upon
multiplication with the result
\be
T_q{\hat T}_{q'}={ A}_{pp'q}^L\btau_\ell(t_q)+{\hat A}_{pp'q}^L
{\mbox{\mycal X}}^{\ell}\btau_{N-\ell}(\omega^\ell t_q),
\label{funtt}\ee
where, after applying (\ref{square3b}) and (\ref{square4b}),
\ba\fl
{\btau_\ell}(t_q)_{\sigma,\sigma'}=\prod_{i=1}^L 
U^{(\ell)}_{pp'q}(\sigma\vp_i,\sigma\vp_{i+1},\sigma'_{i+1},\sigma'_{i})
\nonumber\\
\fl\qquad=\prod_{i=1}^L\sum_{m_i=0}^{\ell-1}
\omega^{\sigma'_i-\sigma\vp_{i+1}}F_{pq}(\ell,\sigma\vp_i-\sigma'_i,m\vp_i)
F_{p'q}(\ell,\sigma\vp_{i+1}-\sigma'_{i+1},m\vp_i)
\frac{\eta_{q,\ell,\sigma\vp_{i}-\sigma'_{i}}}{\eta_{q,\ell,m\vp_i}}.
\label{taul}\ea
Here we have used $\prod_{i=1}^L\eta_{q,\ell,\sigma\vp_{i+1}-\sigma'_{i+1}}
=\prod_{i=1}^L\eta_{q,\ell,\sigma\vp_{i}-\sigma'_{i}}$, which is valid because
of the cyclic boundary condition.
On the other hand the lower diagonal block is given by (\ref{square6}),
where $\ell\le\alpha,\beta\le N-1$, so that $0\le\alpha-\ell,\beta-\ell\le N-1-\ell$. 
By comparing (\ref{square6}) with (\ref{square4b}), we find that
the product of lower diagonal blocks is the $\btau_{N-\ell}(\omega^\ell t_q)$
in (\ref{taul}), except for a shift of the spins $a,b\to a-\ell, b-\ell$ in (\ref{square6}).
Thus the shift operator ${\mbox{\mycal X}}^\ell$ shifting all spins by $\ell$
must be applied to $\btau_{N-\ell}(\omega^\ell t_q)$ in (\ref{funtt}).

Functional relation (\ref{funtt}) is (BBP:3.46) in \cite{BBP}, or (Baxter:3.5)
in \cite{BaxterSB} with a simplification of notation due to Baxter.
Equation (\ref{taul}) is identical to (BBP:3.44a) with $k=0$ and $j=\ell$.
It is easy to verify that (\ref{App'q}) and (\ref{hApp'q}) agree with
(BBP:3.41), (BBP:3.24), (BBP:3.35) and (BBP:3.36).
Comparing  (\ref{App'q}) and (\ref{hApp'q}) with (Baxter:3.5) with its constants
given in (Baxter:2.4) and (Baxter:2.5), we find that the $\btau_\ell$  in \cite{BaxterSB}
is multiplied by a factor $(y_py_{p'})^{(\ell-1)L}$. 
\section{Functional relation in CSOS model}
Choosing $(x\vp_{q'},y\vp_{q'},\mu\vp_{q'})=(y\vp_q,\omega^\ell x\vp_q,\mu_q^{-1})$
and  $(x\vp_{p'},y\vp_{p'},\mu\vp_{p'})=(y\vp_p,\omega^j x\vp_p,\mu_p^{-1})$,
we can easily show that the square $U_{pp'qq'}=0$ for $0\le d-c\le j-1$,
and $j\le a-b\le N-1$. By restricting $n_i=\sigma_i-\sigma_{i+1}$ to be in the
interval $0\le n_i\le j-1$, the $\btau_\ell$-model in (\ref{taul}) becomes the
CSOS model denoted by  $\btau_{\ell,j}$. In this restricted space with
$0\le n_i\le j-1$, the functional relation (\ref{funtt}) still holds, but the constants
in (\ref{omega}), (\ref{App'q}) and (\ref{hApp'q}) are changed according to
\ba
\Omega_{pp'q}\to\Omega_{pq}=\frac{(1-\omega^{-j}y_q/x_{p})(1-x_q/y_{p})}
{(1-y_q^N/x_{p}^N)},\nonumber\\
A_{pp'q}\to A_{pq}=\big(\omega^{\ell-j}t_q/t_{p};\omega\big)^{\vp}_{N-\ell}\left[
\frac{N(1-\omega^{-j}y_q/x_{p})(1-x_q/y_{p})}
{(1-y_q^N/x_{p}^N) (1-x^N_q/y^N_{p})}\right],\nonumber\\
{\hat A}_{pp'q}\to{\hat A}_{pq}=\omega^{(1-j)\ell}({t_q}/{t_{p}};\omega)^{\vp}_{\ell}
\left[\frac{N(1-\omega^{-j}y_q/x_{p})(1-x_q/y_{p})}
{(1-y_q^N/x_{p}^N)(1-x^N_q/y^N_{p})}\right],
\label{Apq}\ea
so that (\ref{funtt}) becomes, 
\ba\fl
T_q{\hat T}_{q'}=\left[\frac{N(1-\omega^{-j}y_q/x_{p})(1-x_q/y_{p})}
{(1-y_q^N/x_{p}^N)(1-x^N_q/y^N_{p})}\right]^L
\nonumber\\
\times\,
\bigg[\big(\omega^{\ell-j}t_q/t_{p};\omega\big)^{L}_{N-\ell}\btau_{\ell,j}(t)
+\omega^{\ell [(1-j)L-Q]}({t_q}/{t_{p}};\omega)^{L}_{\ell}\btau_{N-\ell,j}(\omega^\ell t)\bigg].
\label{funttsos}\ea
From (\ref{taul}) we see that the transfer matrix of the CSOS model is given by
\be
\btau_{\ell,j}(t)=\prod_{i=1}^L U^{(\ell,j)}_{pq}(\sigma\vp_i,\sigma\vp_{i+1},\sigma'_{i+1},\sigma'_{i}).
\label{taulj}\ee
Now the differences of vertical spins are restricted to the values
$0\le \sigma\vp_i-\sigma'_i\le \ell-1$, while the horizontal spin differences are
restricted to  $0\le \sigma\vp_i-\sigma\vp_{i+1},\sigma'_i-\sigma'_{i+1}\le j-1$.
Therefore, from (\ref{square4a}) and (\ref{defFpq}) we find
the weights of the CSOS model to be
\be\fl
U^{(\ell,j)}_{pq}(a,b,c,d)
=\sum_{m=0}^{\ell-1}\omega^{(d-b-j)m}
\Phi\big(1,{\omega^\alpha t}\big)_m^{\alpha,\ell-\alpha-1}
\Phi\big(1,\omega^{m-j}t\big)_{\beta}
^{m,\ell-1-m}.
\label{squaresos}\ee

\subsection{The functional relations}
The $\btau_\ell$ models also satisfy functional relations among themselves,
namely \cite{BBP,BaxterSB}
\ba
\btau_\ell(t)\btau_2(\omega^{\ell-1}t)
={\mbox{\mycal X}}z(\omega^{\ell-1}t)\btau_{\ell-1}(t)+\btau_{\ell+1}(t)
\label{taul2}\ea
with 
\ba
z(t)=[\omega\mu_p\mu_{p'}(t_p-t_q)(t_{p'}-t_q)/(y_py_{p'})^2]^L.
\ea
This is valid for any $p$ and $p'$, meaning that these relations also hold
for the CSOS model. Consequently, we find the functional relation for
the transfer-matrix eigenvalues of the CSOS models to be
\ba\fl
\tau_{\ell,j}(t)\tau_{2,j}(\omega^{\ell-1}t)
=\omega^{-Q+(1-j)L}(1-\omega^{\ell-1}t)^L(1-\omega^{\ell-1-j}t)^L\tau_{\ell-1,j}(t)
+\tau_{\ell+1,j}(t),
\label{funlj}\ea
where we have replaced the shift operator ${\mbox{\mycal X}}$ by its eigenvalue
$\omega^{-Q}$.
Since we have adopted the convention of multiplication from up to down,
the $\Btau2$ matrices here are the transpose of those in \cite{BBP,BaxterSB}.
\subsection{The T-system relations}
To show that the T-system functional relations discussed
in \cite{KlPearce,JKlSuzuki,KNS} also hold for generic $\btau_j$ models,
we may rewrite (\ref{taul2}) as
\be
\tau_{\ell-1}(\omega t)\tau_2(\omega^{\ell-1}t)
=\omega^{-Q}z(\omega^{\ell-1}t)\tau_{\ell-2}(\omega t)+\tau_{\ell}(\omega t).
\label{taul-12}\ee
Multiplying both sides of this equation by $\btau_{\ell}(t)$ and
using (\ref{taul2}), we find
\be\fl 
\omega^{-Q}z(\omega^{\ell-1}t)
[\tau_{\ell-1}(\omega t)\tau_{\ell-1}(t)-\tau_{\ell-2}(\omega t)\tau_{\ell}(t)]
=\tau_{\ell}(\omega t)\tau_{\ell}(t)-\tau_{\ell-1}(\omega t)\tau_{\ell+1}(t).
\label{tautau}\ee
By iteration, we obtain the T-functional relation
\be
\tau_{\ell}(\omega t)\tau_{\ell}(t)-\tau_{\ell-1}(\omega t)\tau_{\ell+1}(t)
=\omega^{-(\ell-1) Q}\prod_{j=1}^{\ell-1}z(\omega^j t).
\label{tauY}\ee

\section{Relation between the coefficients}
As mentioned earlier, there are sixteen quadratic relations between the elements
of the mono\-dromy matrix, with coefficients only depending on the form of the
six-vertex model weights. The first four are $[{\bf A}(x),{\bf A}(y)]=
[{\bf B}(x),{\bf B}(y)]=[{\bf C}(x),{\bf C}(y)]=[{\bf D}(x),{\bf D}(y)]=0$.
Two of the relations are given in (\ref{AB})
and  (\ref{DB}), and two very similar ones are
 \ba
(\omega y-x ){\bf B}(x){\bf A}(y)=
(\omega-1)x{\bf B}(y){\bf A}(x)+(y-x){\bf A}(y){\bf B}(x),
\label{AB2}\\
(\omega y-x ){\bf D}(y){\bf B}(x)=
\omega(y-x ){\bf B}(x){\bf D}(y)+
x(\omega-1){\bf D}(x){\bf B}(y).
\label{DB2}\ea
Adding (\ref{AB}) and (\ref{AB2}), we find
\be
{\bf A}(x){\bf B}(y)+ {\bf B}(x){\bf A}(y)
={\bf B}(y){\bf A}(x)+ {\bf A}(y){\bf B}(x).
\label{AB3}\ee
Now we add (\ref{AB2}) to $x$ times (\ref{AB3}) to find
\be
x[{\bf A}(x){\bf B}(y)- \omega {\bf B}(y){\bf A}(x)]
=y[{\bf A}(y){\bf B}(x)- \omega{\bf B}(x){\bf A}(y)].
\label{AB4}\ee
Expanding and equating the coefficients of $x^m y^\ell$ in
(\ref{AB3}) and (\ref{AB4}), we find
\ba
{\bf B}_m{\bf A}_\ell-{\bf A}_\ell{\bf B}_m
={\bf B}_\ell{\bf A}_m-{\bf A}_m{\bf B}_\ell,
\nonumber\\
{\bf A}_\ell{\bf B}_m-\omega{\bf B}_m{\bf A}_\ell
={\bf A}_{m-1}{\bf B}_{\ell+1}-\omega{\bf B}_{\ell+1}{\bf A}_{m-1}.
\label{ABlm}\ea
Similarly, we may use (\ref{DB}) and (\ref{DB2}) to obtain
\ba
\omega{\bf B}_m{\bf D}_\ell-{\bf D}_\ell{\bf B}_m
=\omega{\bf B}_\ell{\bf D}_m-{\bf D}_m{\bf B}_\ell,
\nonumber\\
{\bf D}_\ell{\bf B}_m-{\bf B}_m{\bf D}_\ell
={\bf D}_{m-1}{\bf B}_{\ell+1}-{\bf B}_{\ell+1}{\bf D}_{m-1}.
\label{BDml}\ea
Since ${\bf B}_0=0$, we find by letting $\ell=0$
\be
{\bf A}_0{\bf B}_m={\bf B}_m{\bf A}_0,\quad
{\bf D}_0{\bf B}_m=\omega{\bf B}_m{\bf D}_0.
\label{A0D0Bm}\ee
Setting $m=1$ in the second equation of (\ref{ABlm}), we find
\be
{\bf A}_\ell{\bf B}_1-\omega{\bf B}_1{\bf A}_\ell
={\bf A}_0{\bf B}_{\ell+1}-\omega{\bf B}_{\ell+1}{\bf A}_0
=(1-\omega){\bf B}_{\ell+1}{\bf A}_0.
\label{AlB1}\ee
From (\ref{BDml}) and (\ref{A0D0Bm}), we also have
\be{\bf D}_m{\bf B}_1-{\bf B}_1{\bf D}_m
={\bf D}_0{\bf B}_{m+1}-{\bf B}_{m+1}{\bf D}_0
=(\omega-1){\bf B}_{m+1}{\bf D}_0.
\label{DmB1}\ee
The next four equations 
 \ba
(\omega y-x ){\bf C}(y){\bf A}(x)=
(\omega-1)x{\bf C}(x){\bf A}(y)+\omega(y-x){\bf A}(x){\bf C}(y),
\nonumber\\
(\omega y-x ){\bf A}(y){\bf C}(x)=
(\omega-1)y{\bf A}(x){\bf C}(y)+(y-x){\bf C}(x){\bf A}(y),
\label{AC}\\
(\omega y-x ){\bf C}(x){\bf D}(y)=
\omega(y-x ){\bf D}(y){\bf C}(x)+
y(\omega-1){\bf C}(y){\bf D}(x),
\nonumber\\
(\omega y-x ){\bf D}(x){\bf C}(y)=
(y-x ){\bf C}(y){\bf D}(x)+
x(\omega-1){\bf D}(y){\bf C}(x),
\label{DC}\ea
yield
\ba
{\bf C}_m{\bf A}_\ell-\omega{\bf A}_\ell{\bf C}_m
={\bf C}_\ell{\bf A}_m-\omega{\bf A}_m{\bf C}_\ell,
\nonumber\\
{\bf A}_\ell{\bf C}_m-{\bf C}_m{\bf A}_\ell
={\bf A}_{m+1}{\bf C}_{\ell-1}-{\bf C}_{\ell-1}{\bf A}_{m+1},
\nonumber\\
{\bf C}_m{\bf D}_\ell-{\bf D}_\ell{\bf C}_m
={\bf C}_\ell{\bf D}_m-{\bf D}_m{\bf C}_\ell,\nonumber\\
\omega{\bf D}_\ell{\bf C}_m-{\bf C}_m{\bf D}_\ell
=\omega{\bf D}_{m+1}{\bf C}_{\ell-1}-{\bf C}_{\ell-1}{\bf D}_{m+1}.
\ea
For the particular value of $m=0$ or $\ell=0$, and using ${\bf C}_{-1}=0$, we find
\ba
{\bf A}_0{\bf C}_m={\bf C}_m{\bf A}_0,\quad \omega{\bf D}_0{\bf C}_m={\bf C}_m{\bf D}_0,\nonumber\\
{\bf C}_0{\bf A}_\ell-\omega{\bf A}_\ell{\bf C}_0=(1-\omega){\bf C}_\ell{\bf A}_0,\nonumber\\
{\bf C}_0{\bf D}_\ell-{\bf D}_\ell{\bf C}_0=(\omega-1){\bf D}_0{\bf C}_\ell.
\label{ACD0}\ea
The remaining four equations are
\ba \fl
( y-x )[{\bf D}(y){\bf A}(x)-{\bf A}(x){\bf D}(y)]=(1-\omega)
[x{\bf C}(x){\bf B}(y)-y{\bf C}(y){\bf B}(x)],\label{ADBC1}
\\ \fl
(y-x )[{\bf D}(x){\bf A}(y)-{\bf A}(y){\bf D}(x)]=(\omega-1)
[x{\bf B}(y){\bf C}(x)-y{\bf B}(x){\bf C}(y)],\label{ADBC2}
\\ \fl
 (y-x )[{\bf C}(y){\bf B}(x)-\omega{\bf B}(x){\bf C}(y)]=(\omega-1)
x[{\bf D}(x){\bf A}(y)-{\bf D}(y){\bf A}(x)],\label{ADBC3}
\\ \fl
( y-x )[{\bf C}(x){\bf B}(y)-\omega{\bf B}(y){\bf C}(x)]=(\omega-1)
y[{\bf A}(y){\bf D}(x)-{\bf A}(x){\bf D}(y)],
\label{ADBC4}\ea
Letting $y\leftrightarrow x$ in (\ref{ADBC2}) and comparing with (\ref{ADBC1})
we  find
\be
x[{\bf C}(x){\bf B}(y)-\omega{\bf B}(y){\bf C}(x)]
=y[{\bf C}(y){\bf B}(x)-\omega{\bf B}(x){\bf C}(y)];
\ee
while letting $y\leftrightarrow x$ in (\ref{ADBC4}) and comparing
with (\ref{ADBC3}) we obtain
\be
[{\bf A}(x){\bf D}(y)-{\bf D}(y){\bf A}(x)]=[{\bf A}(y){\bf D}(x)-{\bf D}(x){\bf A}(y)].
\ee
Consequently, we have
\ba
{\bf A}_m{\bf D}_\ell-{\bf D}_\ell{\bf A}_m
={\bf A}_\ell{\bf D}_m-{\bf D}_m{\bf A}_\ell,
\nonumber\\
{\bf C}_\ell{\bf B}_m-\omega{\bf B}_m{\bf C}_\ell
={\bf C}_{m-1}{\bf B}_{\ell+1}-\omega{\bf B}_{\ell+1}{\bf C}_{m-1}.
\label{CBshift}\ea
From (\ref{ADBC3}), we find
\ba\fl
{\bf C}_\ell{\bf B}_m-\omega{\bf B}_m{\bf C}_\ell
&={\bf C}_{\ell-1}{\bf B}_{m+1}
-\omega{\bf B}_{m+1}{\bf C}_{\ell-1}+
(1-\omega)[{\bf D}_m{\bf A}_\ell-{\bf D}_\ell{\bf A}_m],
\nonumber\\
\fl
&={\bf C}_{m}{\bf B}_{\ell}-\omega{\bf B}_{\ell}{\bf C}_{m}+
(1-\omega)[{\bf D}_m{\bf A}_\ell-{\bf D}_\ell{\bf A}_m].
\ea
Since ${\bf B}_0=0$, we find
\be
{\bf C}_0{\bf B}_m-\omega{\bf B}_m{\bf C}_0
=(1-\omega)[{\bf D}_m{\bf A}_0-{\bf D}_0{\bf A}_m].
\label{C0Bm}\ee
Using ${\bf C}_L=0$, we obtain
\be
{\bf C}_\ell{\bf B}_L-\omega{\bf B}_L{\bf C}_\ell
=(1-\omega)[{\bf D}_L{\bf A}_\ell-{\bf D}_\ell{\bf A}_L].
\label{CBL}\ee
Particularly, we find from (\ref{C0Bm}) and (\ref{CBshift}) that
\ba\fl
{\bf C}_0{\bf B}_L-\omega{\bf B}_L{\bf C}_0
={\bf C}_{L-1}{\bf B}_1-\omega{\bf B}_1{\bf C}_{L-1}
=(1-\omega)[{\bf D}_L{\bf A}_0-{\bf D}_0{\bf A}_L].
\label{CB0L}\ea
Putting $m=1$ in (\ref{C0Bm}), we obtain
\be
{\bf C}_0{\bf B}_1-\omega{\bf B}_1{\bf C}_0
=(1-\omega)[{\bf D}_1{\bf A}_0-{\bf D}_0{\bf A}_1],
\label{C0B1}\ee
Setting $\ell=L-1$ in (\ref{CBL}), we obtain
\be
{\bf C}_{L-1}{\bf B}_L-\omega{\bf B}_L{\bf C}_{L-1}
=(1-\omega)[{\bf D}_L{\bf A}_{L-1}-{\bf D}_{L-1}{\bf A}_L].
\label{CL-1BL}\ee
\subsection{Modified Serre relation}
We shall now show the modified Serre relation
\be
\Psi={\bf C}^{(3)}_0{\bf B}\vp_1
-{\bf C}^{(2)}_0{\bf B}\vp_1{\bf C}\vp_0
+\omega{\bf C}\vp_0{\bf B}\vp_1{\bf C}^{(2)}_0
-\omega^3{\bf B}\vp_1{\bf C}^{(3)}_0=0.
\label{serremd1}\ee
We first may rewrite $\Psi$, and then use (\ref{C0B1}) and
$[3]=1+\omega+\omega^2$, to find
\ba\fl
\Psi={\bf C}^2_0[{\bf C}_0{\bf B}_1-\omega{\bf B}_1{\bf C}_0]
+(\omega-[3]){\bf C}_0[{\bf C}_0{\bf B}_1-\omega{\bf B}_1{\bf C}_0]{\bf C}_0
\nonumber\\
-\omega^2[{\bf C}_0{\bf B}_1-\omega{\bf B}_1{\bf C}_0]{\bf C}^{2}_0
=(1-\omega)[{\bf C}_0{\bf H}-\omega^2{\bf H}{\bf C}_0],
\label{psi}\ea
where
\be
{\bf H}={\bf C}_0({\bf D}_1{\bf A}_0-{\bf D}_0{\bf A}_1)
-({\bf D}_1{\bf A}_0-{\bf D}_0{\bf A}_1){\bf C}_0.
\ee
Now we use (\ref{ACD0}) to find
\ba\fl
{\bf H}=({\bf C}_0{\bf D}_1-{\bf D}_1{\bf C}_0){\bf A}_0
+{\bf D}_0({\bf A}_1{\bf C}_0-\omega{\bf C}_0{\bf A}_1)
\nonumber\\
=(\omega^2-1){\bf D}_0({\bf A}_0{\bf C}_1-{\bf A}_1{\bf C}_0).
\label{bH}\ea
Substituting (\ref{bH}) into (\ref{psi}) and using (\ref{ACD0}),
we prove (\ref{serremd1}), as
\ba\fl
\Psi=(1-\omega)(\omega^2-1)\omega{\bf D}_0
[(1-\omega){\bf A}_0{\bf C}_1{\bf C}_0
-({\bf C}_0{\bf A}_1-\omega{\bf A}_1{\bf C}_0){\bf C}_0]=0.
\ea
Similarly, we can show
\ba
{\bf C}\vp_0{\bf B}^{(3)}_1
-{\bf B}\vp_1{\bf C}\vp_0{\bf B}^{(2)}_1
+\omega{\bf B}^{(2)}_1{\bf C}\vp_0{\bf B}\vp_1
-\omega^3{\bf B}^{(3)}_1{\bf C}\vp_0=0,
\nonumber\\
{\bf C}^{(3)}_{L-1}{\bf B}\vp_L
-{\bf C}^{(2)}_{L-1}{\bf B}\vp_L{\bf C}\vp_{L-1}
+\omega{\bf C}\vp_{L-1}{\bf B}\vp_L{\bf C}^{(2)}_{L-1}
-\omega^3{\bf B}\vp_L{\bf C}^{(3)}_{L-1}=0,
\nonumber\\
{\bf C}\vp_{L-1}{\bf B}^{(3)}_L
-{\bf B}\vp_L{\bf C}\vp_{L-1}{\bf B}^{(2)}_L
+\omega{\bf B}^{(2)}_L{\bf C}\vp_{L-1}{\bf B}\vp_L
-\omega^3{\bf B}^{(3)}_L{\bf C}\vp_{L-1}=0.
\label{serremd2}\ea

\end{document}